\documentclass[11pt]{article}
\usepackage{pepadun}
\usepackage{xcolor}

\usepackage{bbding} 
\usepackage{tabularx}
\usepackage{svg}

\usepackage{graphicx}
\usepackage{subcaption} 

\usepackage{hyperref}
\usepackage{fontawesome}

\usepackage{fancyhdr}
\usepackage{lipsum}

\usepackage{cite}

\makeatletter
\def\blfootnote{\xdef\@thefnmark{}\@footnotetext}
\makeatother

\newcommand{\module}[1]{\textbf{#1}}
\newcommand{\plugin}[1]{\textit{#1}}
\newcommand{\class}[1]{\texttt{#1}}
\newcommand{\method}[1]{\texttt{#1}}  
\newcommand{\annotation}[1]{\texttt{@#1}}
\newcommand{\code}[1]{\texttt{#1}}

\title{\raisebox{-0.35em}{\includegraphics[height=1.5em]{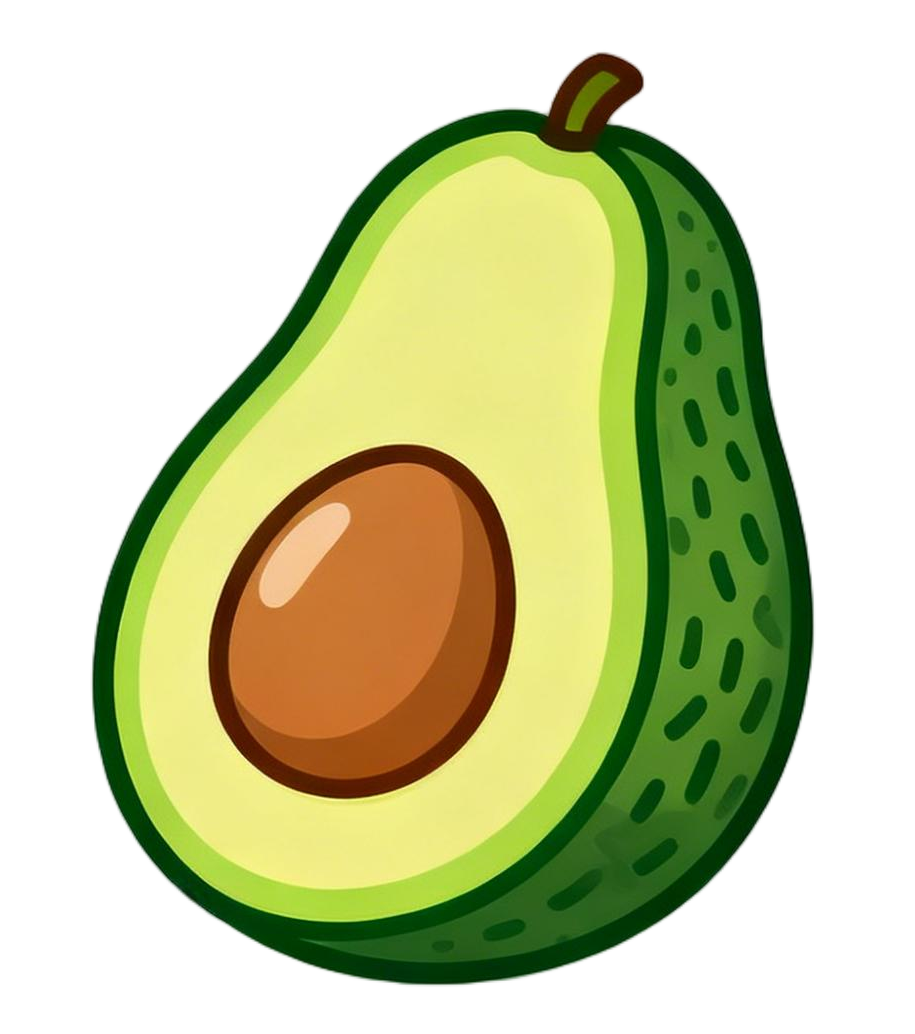}} Agent-Kernel: A MicroKernel Multi-Agent System Framework for Adaptive Social Simulation Powered by LLMs}

\author{%
\hspace*{-0.8cm}
    \begin{tabular}{c}
        \textsuperscript{1,2}Yuren Mao, 
        \textsuperscript{1}Peigen Liu, 
        \textsuperscript{1}Xinjian Wang, 
        \textsuperscript{1}Rui Ding, 
        \textsuperscript{1}Jing Miao, 
        \textsuperscript{1}Hui Zou, 
        \textsuperscript{1}Mingjie Qi, 
        \textsuperscript{1}Wanxiang Luo, \\
        \textsuperscript{4}Longbin Lai, 
        \textsuperscript{6}Kai Wang, 
        \textsuperscript{5}Zhengping Qian, 
        \textsuperscript{3}Peilun Yang, 
        \textsuperscript{1,2}Yunjun Gao\textsuperscript{*}, 
        \textsuperscript{3}Ying Zhang\textsuperscript{*} \\[6pt]
        \textsuperscript{1}School of Software Technology, Zhejiang University  \\
        \textsuperscript{2}The Key Lab of Big Data Intelligent Computing of Zhejiang Province\\
        \textsuperscript{3}Laboratory for Statistical Monitoring and Intelligent Governance of Common Prosperity, Zhejiang Gongshang University  \\
        \textsuperscript{4}Tongyi Lab, Alibaba Group,        \textsuperscript{5}Alibaba Cloud,
        \textsuperscript{6}Antai College of Economics and Management, Shanghai Jiao Tong University\\
        \end{tabular}
    \begin{tabular}{c}
        \raisebox{-0.25em}{\includegraphics[height=1.2em]{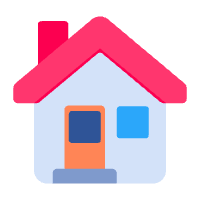}} Homepage:  \href{https://www.agent-kernel.tech}{https://www.agent-kernel.tech}, \\
        \faGithub\ Github: \href{https://github.com/ZJU-LLMs/Agent-Kernel}{https://github.com/ZJU-LLMs/Agent-Kernel}
    \end{tabular}%
}
\date{} 

\begin{document}

\maketitle

\begin{abstract}
Multi-Agent System (MAS) developing frameworks serve as the foundational infrastructure for social simulations powered by Large Language Models (LLMs). However, existing frameworks fail to adequately support large-scale simulation development due to inherent limitations in adaptability, configurability, reliability, and code reusability. For example, they cannot simulate a society where the agent population and profiles change over time. To fill this gap, we propose Agent-Kernel, a framework built upon a novel society-centric modular microkernel architecture. It decouples core system functions from simulation logic and separates cognitive processes from physical environments and action execution. Consequently, Agent-Kernel achieves superior adaptability, configurability, reliability, and reusability. We validate the framework's superiority through two distinct applications: a simulation of the \textit{Universe 25} (Mouse Utopia) experiment, which demonstrates the handling of rapid population dynamics from birth to death; and a large-scale simulation of the \textit{Zhejiang University Campus Life}, successfully coordinating 10,000 heterogeneous agents, including students and faculty.

\end{abstract}


\blfootnote{*Corresponding Author}

\section{Introduction}
\label{sec:introduction}

Large Language Model (LLM)-based social simulation, which exploits the role-playing capabilities of LLMs to simulate human behavior and societal dynamics, has the potential to accelerate social science research and unveil new investigative opportunities \cite{anthis2025llm}. To implement such simulations, adopting an LLM-based Multi-Agent System (MAS), wherein agents assume social roles and interact with one another, is a natural choice. However, developing a MAS specifically for realistic social simulation is a complex software engineering task that imposes distinct requirements beyond those of typical multi-agent applications. We identify four critical dimensions that a MAS framework must address to enable realistic and adaptive social simulation.

\begin{enumerate} \item \textbf{Adaptability}: Societies are inherently dynamic and evolve over time, characterized by fluctuating populations, evolving actions, and ever-changing environments. Consequently, a MAS framework must support the runtime update of agents, actions, and environments to dynamically accommodate these societal shifts. Furthermore, the framework must be designed for scalability to accommodate such simulations at a large scale.

\item \textbf{Configurability}: Social analysis frequently necessitates the dynamic adjustment of experimental configurations (e.g., spatial layouts, social rules, or agent capabilities) to systematically observe the resulting behavioral changes. Therefore, a MAS framework must support global and dynamic configuration capabilities to facilitate such experimental flexibility.

\item \textbf{Reliability}: Societies are vulnerable to erroneous actions; a single deviation can trigger cascading effects that reshape the societal structure, dynamics, or collective trajectory. To mitigate this, a MAS framework must be equipped with the capability to monitor and verify agent behaviors.

\item \textbf{Reusability}: Given the significant structural and functional diversity of societies across different contexts, a MAS framework should be designed to maximize the reuse of data and code for agents, environments, and actions. This facilitates the simulation of new societies by requiring only minor modifications to existing ones.
\end{enumerate}
\vspace{-0.3cm}

\begin{figure}[tbp]
  \centering
  \vspace{-0.5cm}
\includegraphics[width=0.9\textwidth]{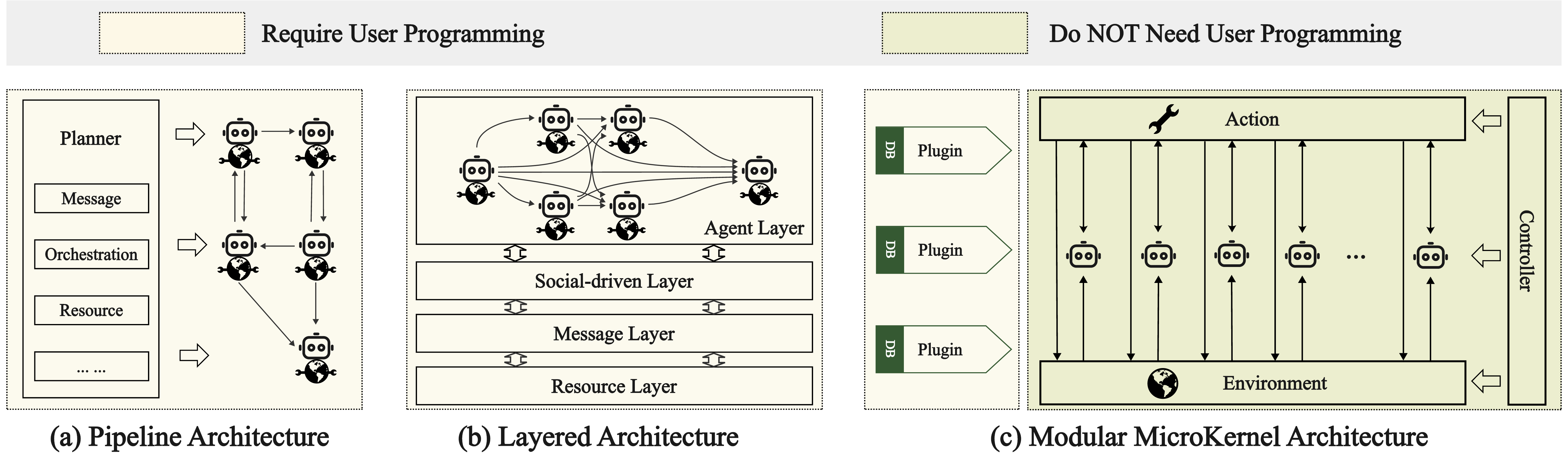}
  \small
  \caption{Comparison between pipeline architecture, layered architecture  and modular microkernel architecture.
  The \raisebox{-0.5em}{\includegraphics[height=1.75em]{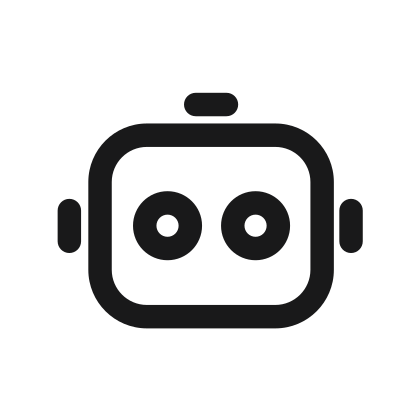}}, 
  \raisebox{-0.35em}{\includegraphics[height=1.4em]{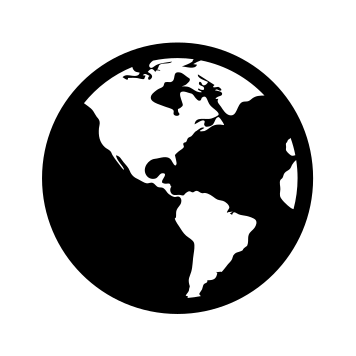}}, and 
  \raisebox{-0.3em}{\includegraphics[height=1.3em]{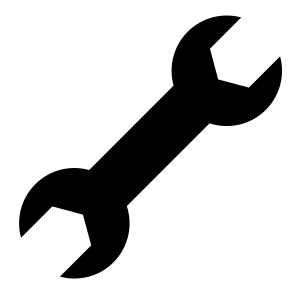}} symbols represent the agent, environment and action module, respectively; and \raisebox{-0.3em}{\includegraphics[height=1.3em]{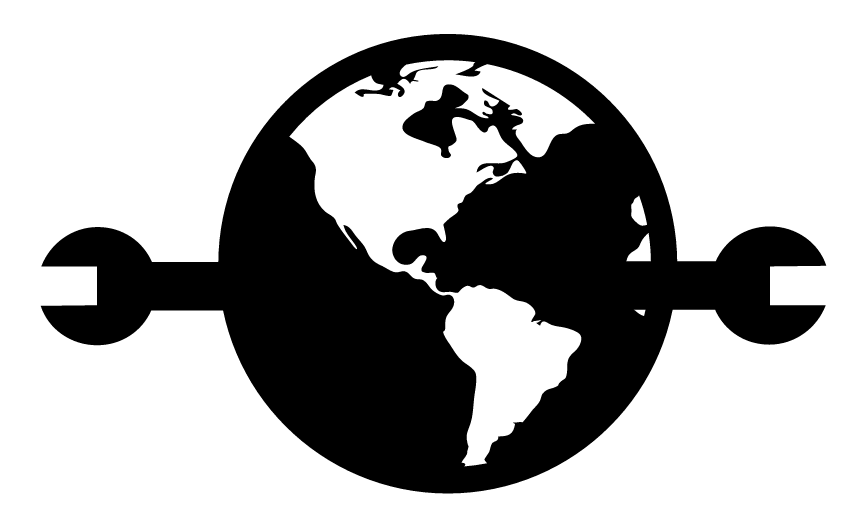}} stands for the coupling of the environment and action modules under the agent.}
  \label{fig:frameworks-comp}
  \vspace{-0.2cm}
\end{figure}

Despite the proposal of numerous MAS development frameworks—such as AgentSociety~\cite{agentsociety2025}, AgentScope~\cite{agentscope2025}, CAMEL~\cite{camel2023}, and CrewAI~\cite{crewai}—current software architectures remain insufficient for meeting the distinct requirements of realistic social simulation. In general, the architectures of these frameworks can be categorized into two paradigms: pipeline architecture (e.g. AgentScope~\cite{agentscope2025}, AutoGen~\cite{autogen2023}) and layered architecture (e.g. AgentSociety~\cite{agentsociety2025}, CAMEL~\cite{camel2023}),
as demonstrated in Fig.\ref{fig:frameworks-comp} (a) and (b) respectively.
 The pipeline architecture employs a centralized planner to allocate resources, manage message flow, and orchestrate agent behaviors according to predefined specifications.
In comparison, the layered architecture decomposes a MAS application into horizontal layers, typically comprising the agent, social-driven, message, and database layers. Crucially, neither the layered nor pipeline architectures satisfies the adaptability requirement. In both approaches, the quantity and profiles of agents are static, as they are pre-defined within the planner of the pipeline architecture and the social-driven layer of the layered architecture, respectively. Furthermore, environment and action parameters are often tightly coupled with agent profiles, which complicates the runtime updating of agents, actions, and the environment. For instance, the introduction of a single new agent necessitates a cascading update of the social relationship lists for all existing agents that might interact with it. This structural rigidity fundamentally hinders the adaptability required for realistic social simulation.
Moreover, these architectures demonstrate significant deficiencies in configurability and reliability, primarily due to the absence of a global control module capable of supporting runtime configuration updates and behavioral verification. Finally, reusability is constrained; both the planner logic and layer implementations in layered architectures necessitate extensive user programming, thereby limiting the potential for code reuse across different applications.

Existing pipeline and layered architectures fail to satisfy these requirements, primarily due to their reliance on an \textbf{agent-centric} design philosophy, that is, to anthropomorphize agents, explicitly coupling them with specific action capabilities and environmental information. However, a fundamental distinction is often overlooked: LLMs function purely as cognitive engines and are devoid of inherent physical constraints. This characteristic presents a unique opportunity to decouple cognitive processes from physical environments and action execution, thereby eliminating the need to embed extensive non-cognitive data directly within the agent. Consequently, it enables a \textbf{society-centric} simulation paradigm. In contrast to the agent-centric view where society emerges as the collectives of agents, this paradigm explicitly models the societal infrastructure—environments, rules, and actions—as independent entities, allowing the simulation to be designed and controlled from a system-level perspective.

Building on this observation, 
we introduce Agent-Kernel, a society-centric modular microkernel architecture consisting of a foundational core and an extensible suite of application-specific plugins, shown as Fig \ref{fig:frameworks-comp} (c). This design effectively decouples the general-purpose simulation infrastructure from application-specific logic. By providing a pre-implemented core system, it allows users to concentrate exclusively on developing plugins.
The core system comprises five distinct modules: \module{Agent}, \module{Environment}, \module{Action}, \module{Controller}, and \module{System}. By decoupling the \module{Agent}, \module{Environment}, and \module{Action} entities, the framework supports efficient, independent runtime updates, thereby achieving high adaptability. Furthermore, we encapsulate specific functionalities of these entities into plugins; this approach promotes reusability by enabling users to develop complex behaviors with minimal code modifications. Finally, the \module{Controller} module serves as the central nexus connecting these decoupled modules and plugins, enabling holistic governance and configuration of the simulation, which ensures reliability and configurability.

\begin{figure}[tbp]
    \centering
    \vspace{-0.5cm}
    \begin{subfigure}[b]{0.45\textwidth} 
        \centering
        \includegraphics[width=\textwidth]{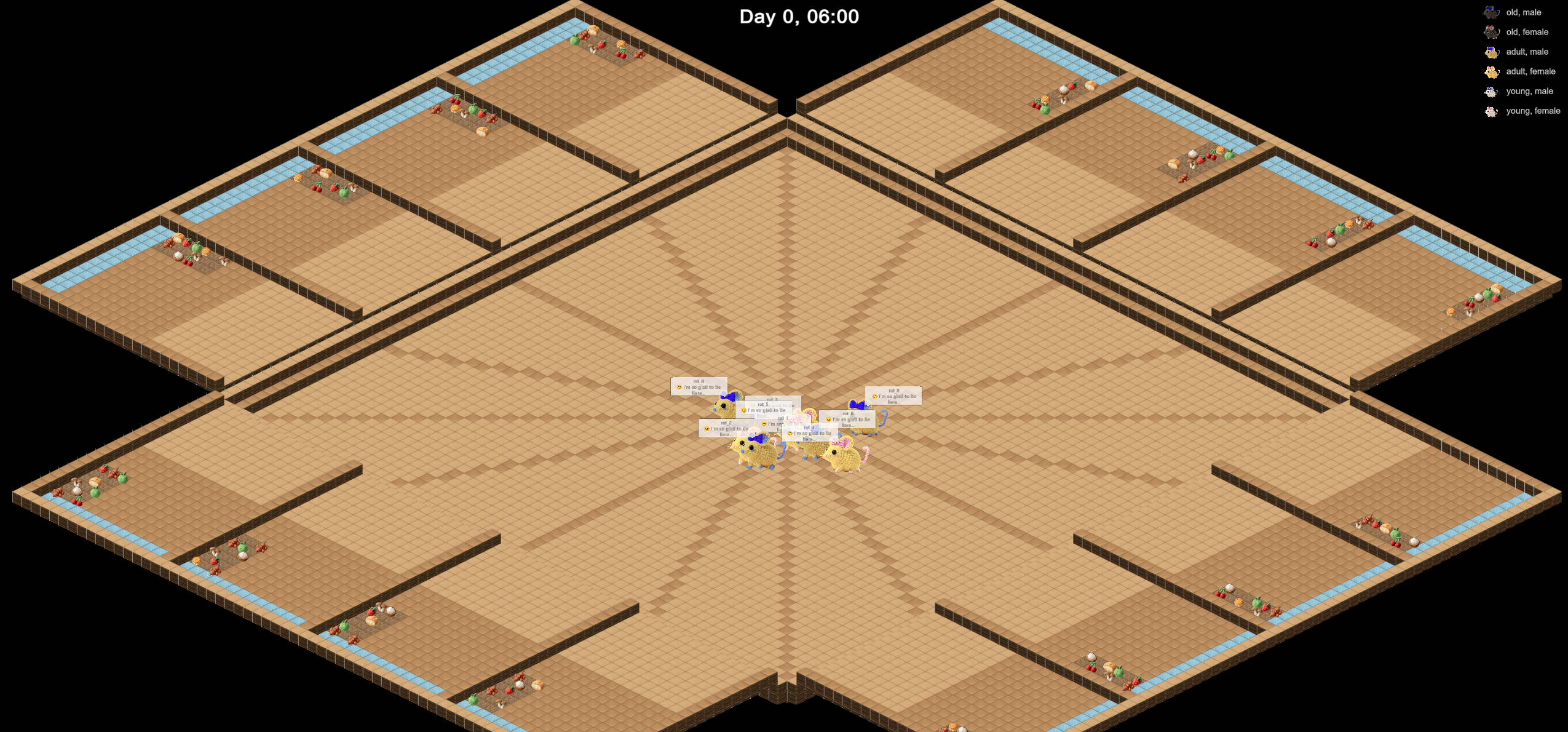} 
        \caption{Initialization.}
        \label{subfig:top_left}
    \end{subfigure}
    \hfill 
    \begin{subfigure}[b]{0.45\textwidth}
        \centering
        \includegraphics[width=\textwidth]{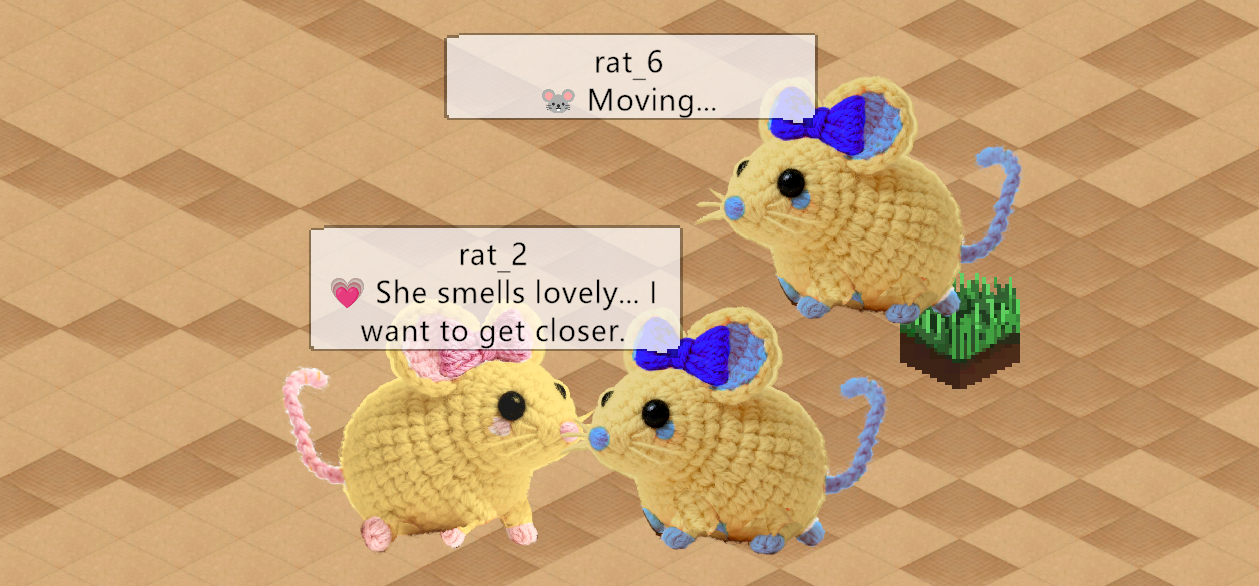}
        \caption{Courtship.}
        \label{subfig:top_right}
    \end{subfigure}
    
    \vspace{1em} 
    \vspace{-0.1cm}
    \begin{subfigure}[b]{0.45\textwidth}
        \centering
        \includegraphics[width=\textwidth]{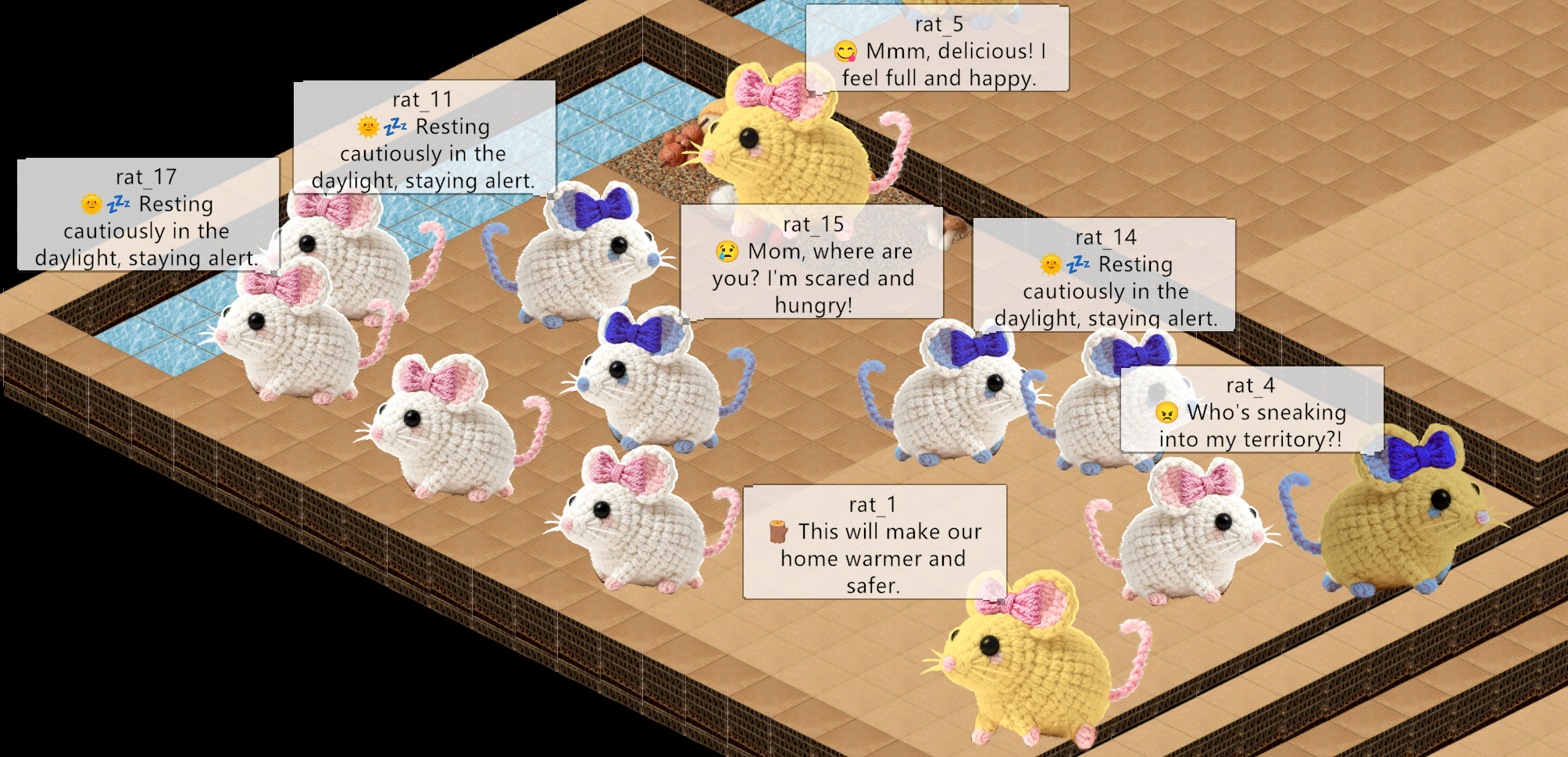}
        \caption{Collective behaviors.}
        \label{subfig:bottom_left}
    \end{subfigure}
    \hfill
    \begin{subfigure}[b]{0.45\textwidth}
        \centering
        \includegraphics[width=\textwidth]{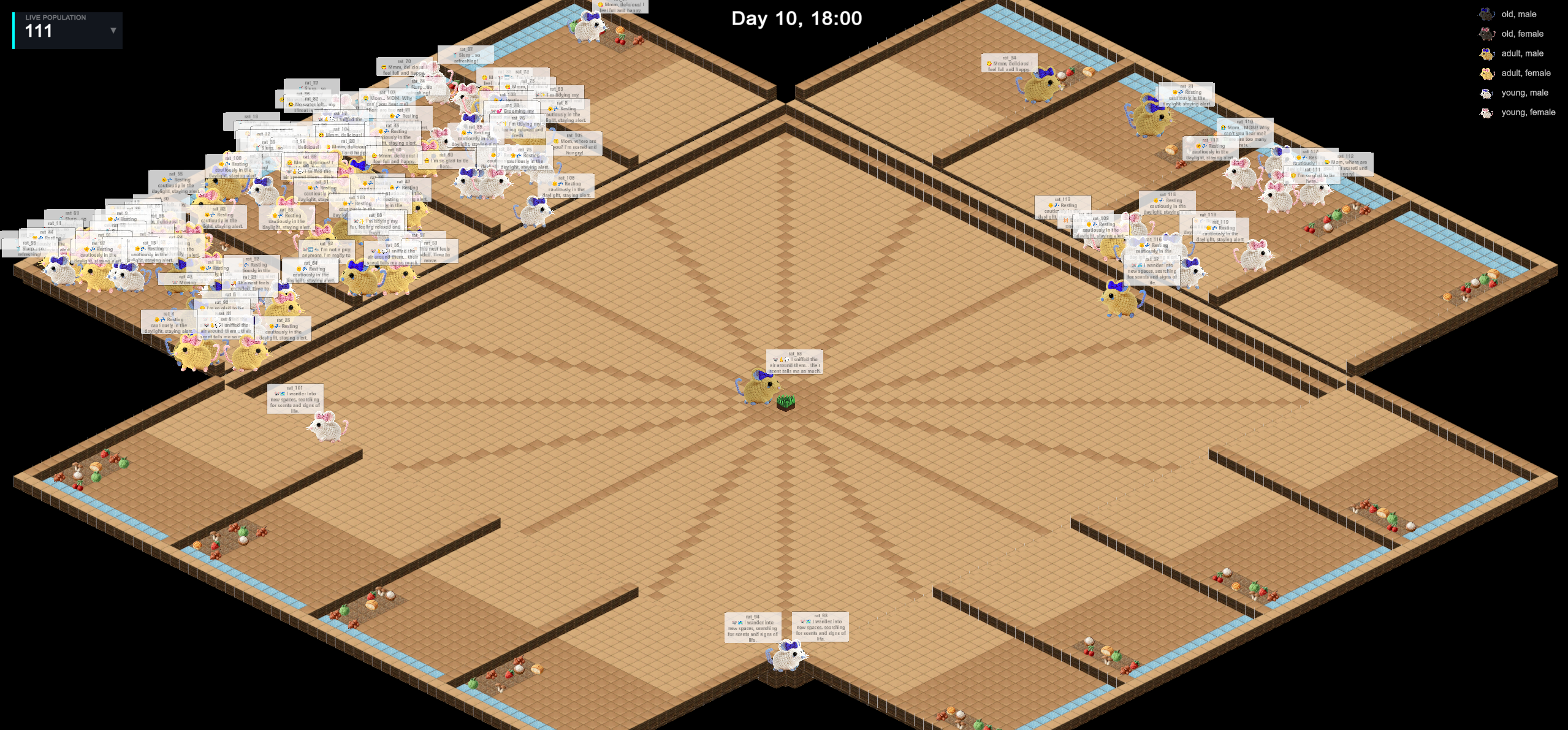}
        \caption{Population growth.}
        \label{subfig:bottom_right}
    \end{subfigure}
\vspace{-0.1cm}
    \small
    \caption{Simulation of the Universe 25 experiment. (a) The initialization stage, featuring an equal sex ratio of four pairs (4 males, 4 females). (b) Mating behavior, showing the male mouse pursuing the female. (c) A representative scenario of collective behaviors. (d) The population growing to 111 individuals.}
    \label{fig:universe25}
    \vspace{-0.2cm}
\end{figure}

\begin{figure}[tbp]
    \centering
    \vspace{-0.5cm}
    \begin{subfigure}[b]{0.45\textwidth}
        \centering
        \includegraphics[width=\linewidth, height=4.0cm]{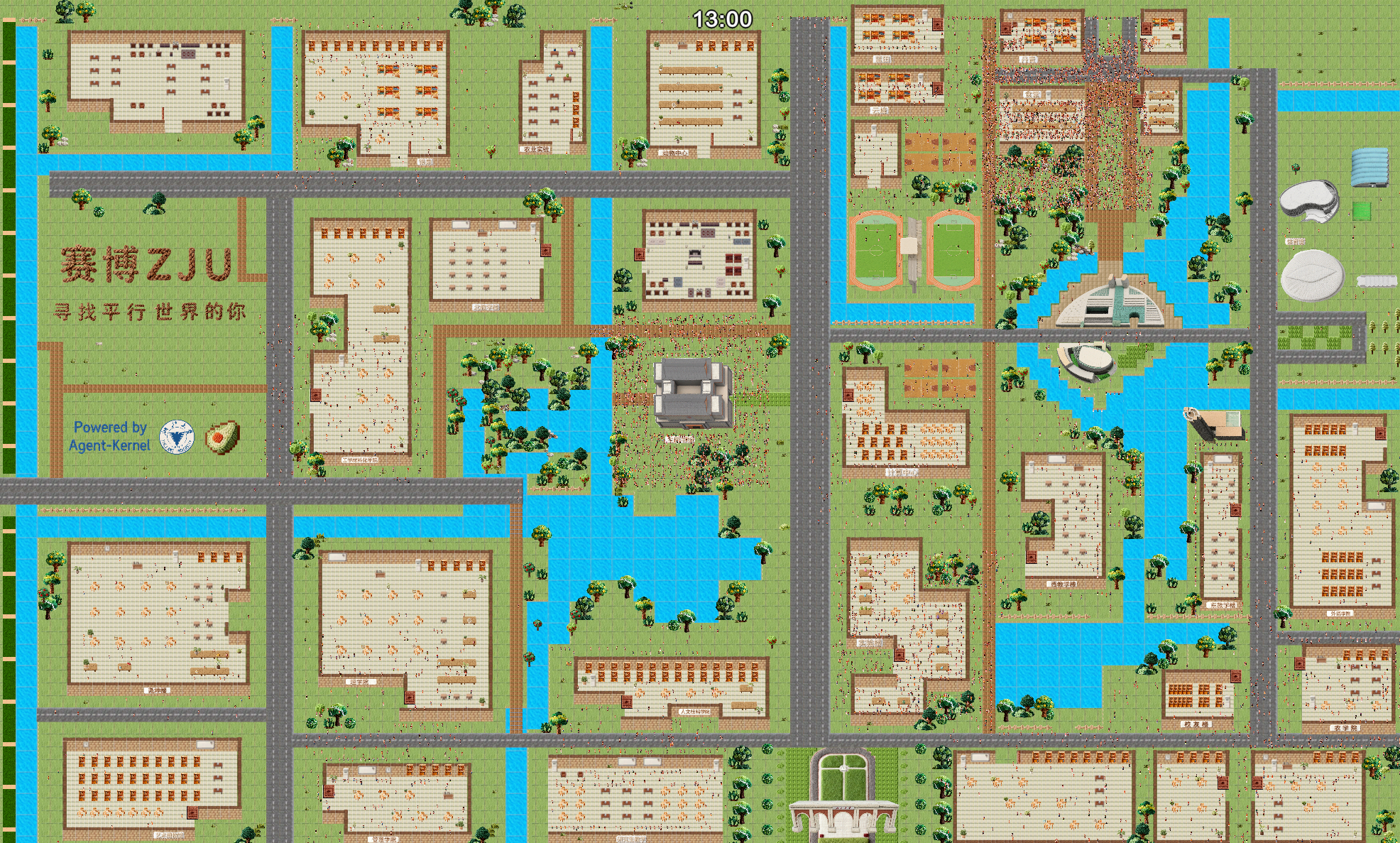} 
        \caption{Panoramic view of the campus.}
    \end{subfigure}
    \hfill 
    \begin{subfigure}[b]{0.45\textwidth}
        \centering
        \includegraphics[width=\linewidth, height=4.0cm]{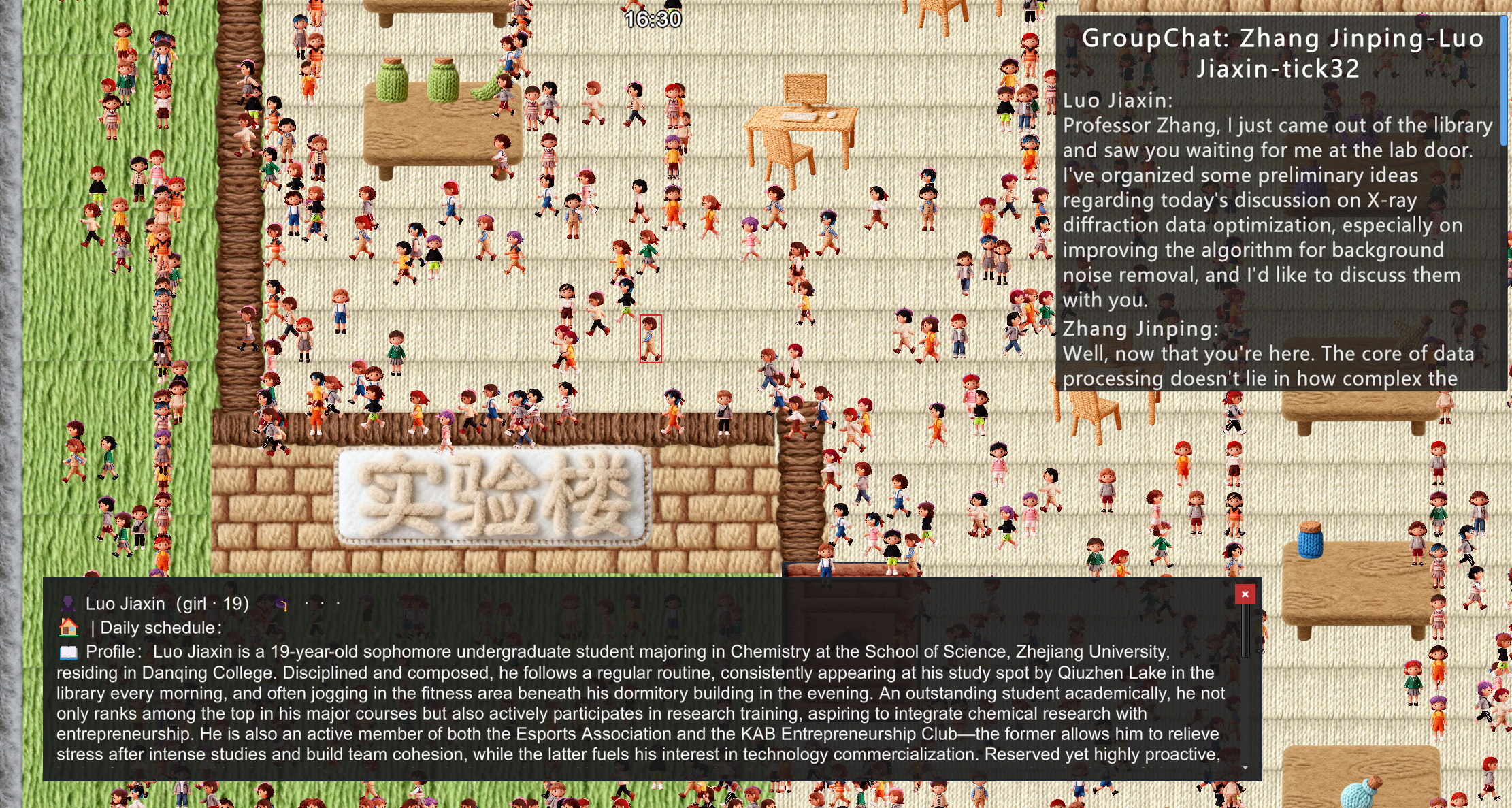}
        \caption{Discussion in a Laboratory.}
    \end{subfigure}
    
    \vspace{1em} 
    \vspace{-0.1cm}
    \begin{subfigure}[b]{0.45\textwidth}
        \centering
        \includegraphics[width=\linewidth, height=4.0cm]{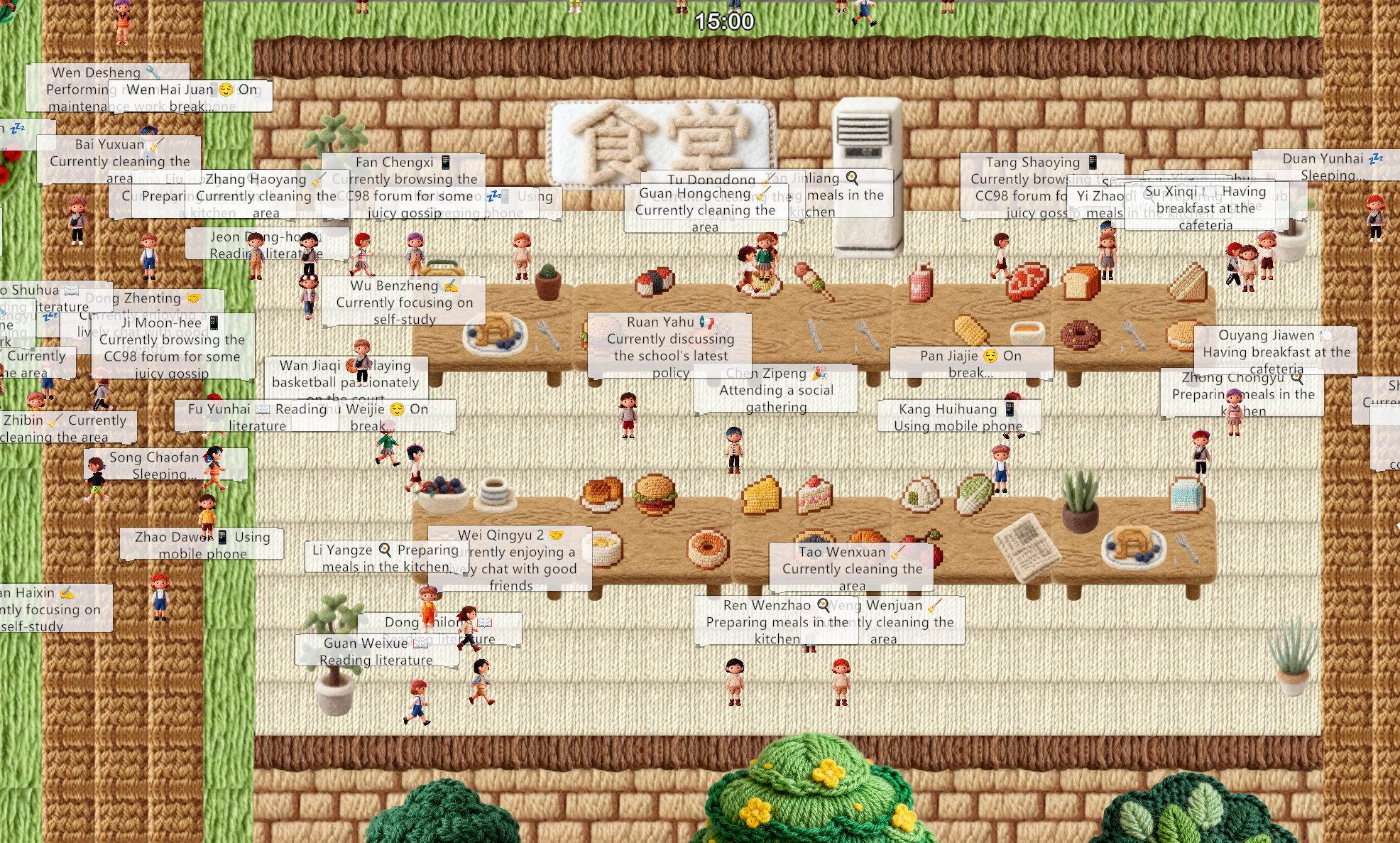}
        \caption{Dining scenario in a canteen.}
    \end{subfigure}
    \hfill
    \begin{subfigure}[b]{0.45\textwidth}
        \centering
        \includegraphics[width=\linewidth, height=4.0cm]{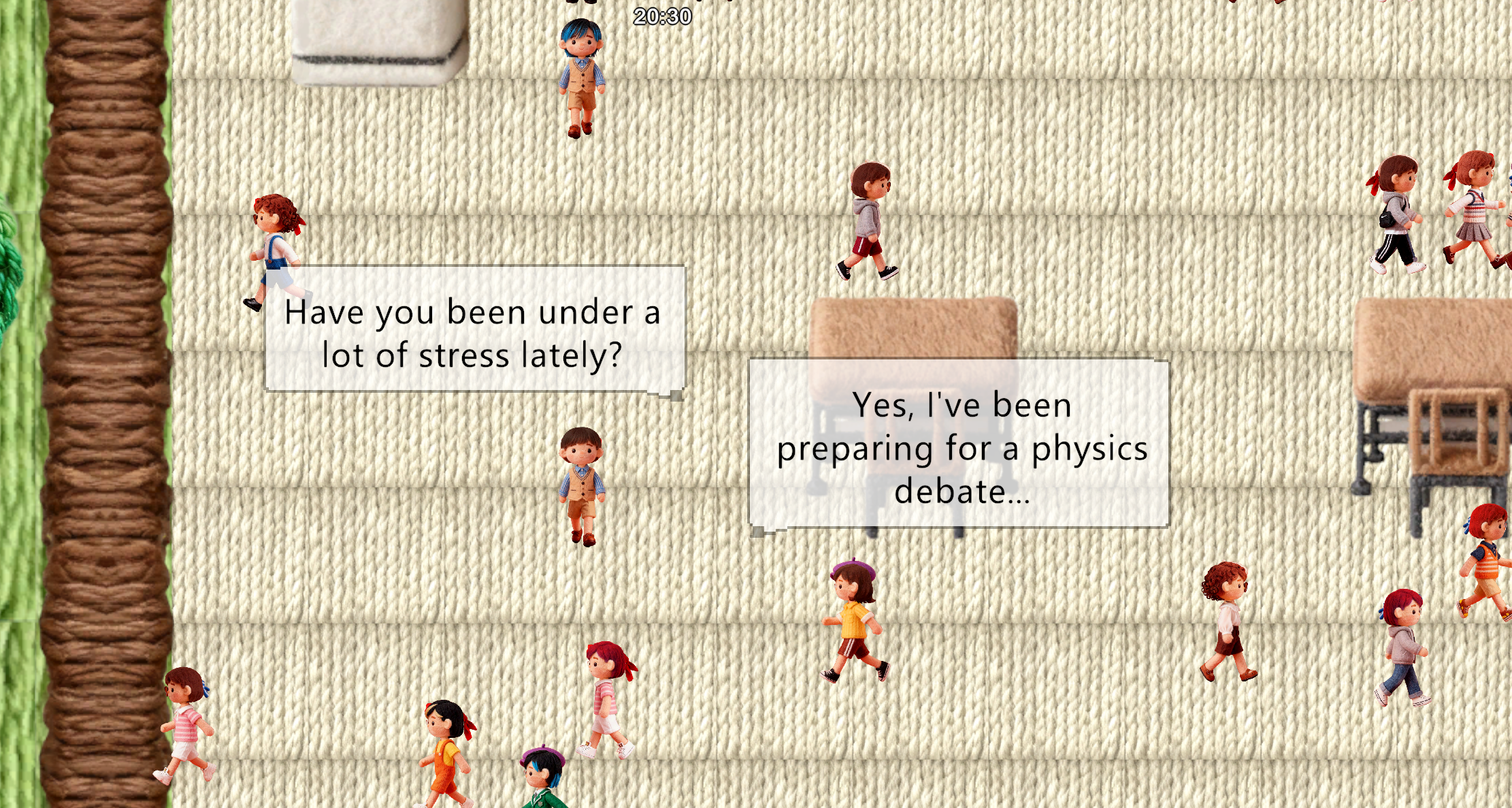}
        \caption{Conversation between two agents.}
    \end{subfigure}
    
    \small
    \caption{Simulation of the Zijingang campus, Zhejiang University. (a) Panoramic view of the campus. (b) Several agents discussing in a laboratory. (c) The dining scenario in a canteen. (d) Conversation between two agents.}
    \label{fig:zjucampus}
    \vspace{-0.1cm}
\end{figure}

To validate the capabilities of Agent-Kernel, we have developed two distinct social simulation applications: the Universe 25 experiment and ZJU Campus Life, illustrated in Fig. \ref{fig:universe25}  and Fig. \ref{fig:zjucampus}  respectively. The Universe 25 scenario simulates the conditions of the historical behavioral sink experiment \cite{calhoun1973}, wherein the agent population scales from an initial cohort of 8 to over 300; this fluctuation, alongside dynamically configurable experimental parameters, validates the framework's adaptability and configurability. Conversely, the ZJU Campus Life scenario models the large-scale environment of the Zijingang campus of Zhejiang University, supporting a population of 10,000 agents acting as students and faculty. Within this simulation, agents engage in complex, heterogeneous behaviors—including mobility, academic research, resource consumption, and social networking—while their profiles and environmental variables remain adaptively mutable and their outputs are regulated by the central controller. Collectively, these applications demonstrate the adaptability, configurability, and reliability of our framework, while the successful sharing of core systems and functional plugins across disparate applications attests to the system's modularity and ease of development.

\section{Overview of Agent-Kernel Framework}

\subsection{Modular Microkernel Architecture}
The Agent-Kernel framework adopts a modular microkernel architecture, comprised of a modular core system and an extensible suite of plugins. The core system serves as the backbone of the Multi-Agent System (MAS), orchestrating essential services including plugin registration, behavior verification, log management,  asynchronous communication, and so on. Conversely, plugins are responsible for implementing the specific functionalities required to simulate a specific society. Each module provides component interfaces that enable plugins to bind seamlessly to the core system.
This microkernel approach effectively decouples general-purpose, system-level MAS programming from the application-specific logic required for social simulation. Consequently, researchers can focus on crafting application-specific functionalities, relying on the pre-developed core system. Furthermore, plugins designed for specific functionalities can be reused across diverse scenarios.


As illustrated in Fig. \ref{fig:framework}, the modular core system consists of five distinct modules: \module{Agent}, \module{Environment}, \module{Action}, \module{Controller}, and \module{System}. 
The \module{Agent}, \module{Environment}, and \module{Action} entities are decoupled into separate modules that collaborate via message passing. This decoupling facilitates the independent modification of agents, actions, and environments—allowing for the adjustment of individual agents—thereby enhancing adaptability. 
Critically, all message passing, whether between agents or modules, is routed through the \module{Controller}. This centralized routing enables the \module{Controller} to validate action feasibility, thus enhancing reliability. 
The \module{Agent}, \module{Environment}, and \module{Action} modules each manage a set of plugins that define application-specific functionalities. These plugins form a shared repository adaptable to different scenarios, offering excellent reusability. 
Additionally, the plugins expose interfaces to the \module{Controller} to facilitate runtime interventions, ensuring the system’s {configurability. The design details of the modular core system are presented in Section \ref{sec:agent_kernel}.


Regarding the plugin architecture, the \module{Agent} module governs agent-related plugins—such as the \plugin{profile}, \plugin{perceive}, \plugin{plan}, \plugin{invoke}, \plugin{state}, and \plugin{reflect} plugins—which shape agent profiles and cognitive processes. These plugins integrate with the \module{Agent} module via their corresponding components (e.g., the \plugin{profile}, \plugin{perceive}, and \plugin{plan} components). Similarly, the \module{Environment} module manages environment-related plugins (e.g., \plugin{space} and \plugin{relation} plugins) utilized to design the physical attributes of the simulated society, such as movement routes and social networks. These link to the \module{Environment} module via their respective components. Finally, the \module{Action} module administers action-related plugins (e.g., \plugin{communication} and \plugin{tools} plugins) that provide feasible agent actions. These plugins connect to the \module{Action} module through their corresponding components. Plugin specifics are elaborated in Section \ref{sec:plugin}, while software implementation details are provided in Section \ref{sec:implementation}.

\begin{figure}[tbp]
    \centering
        \vspace{-0.3cm}
    \includegraphics[width=1\linewidth]{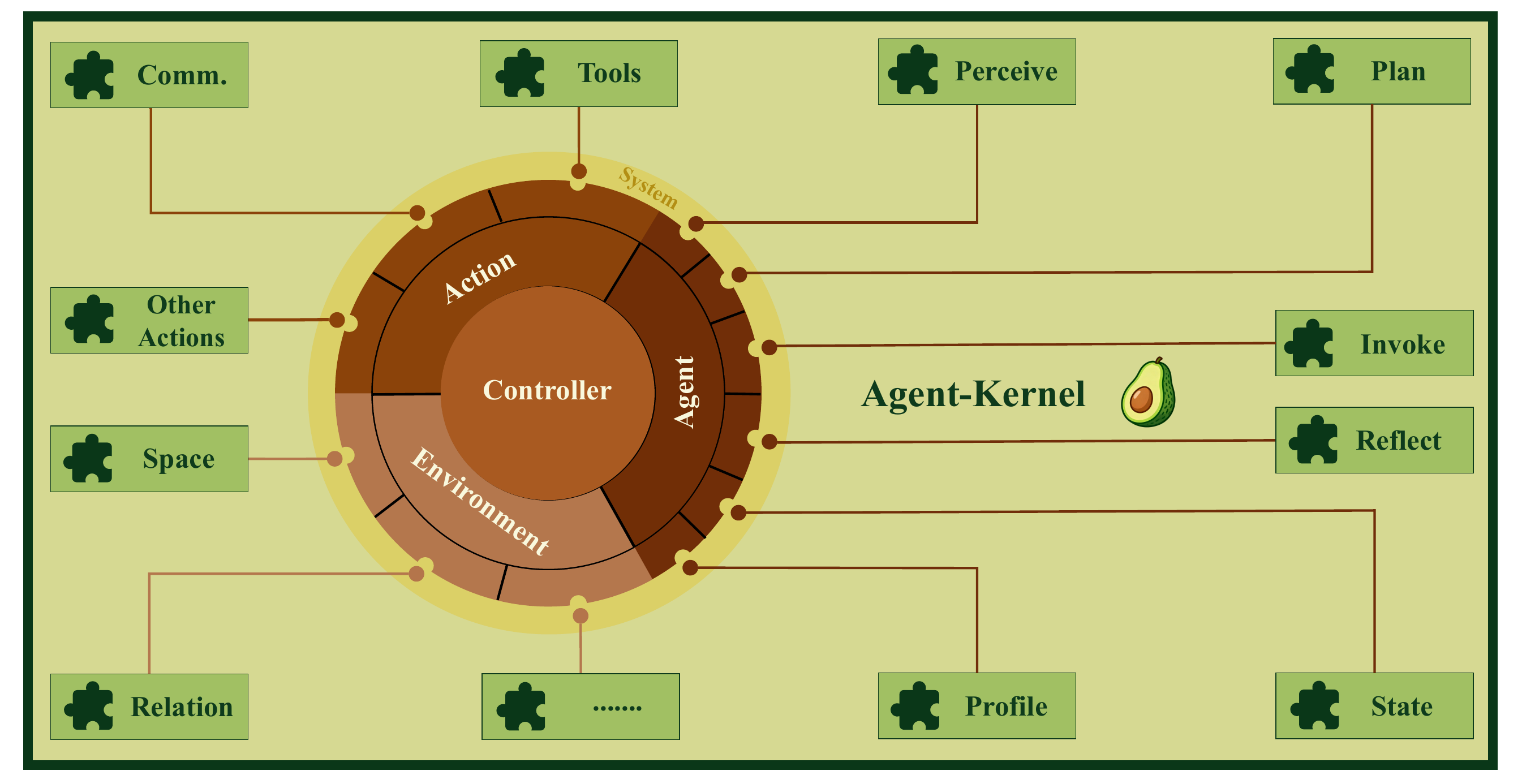}
    
    \caption{Illustration of the  modular microkernel architecture of Agent-Kernel. 
        \raisebox{-0.1em}{\includegraphics[height=1.3em]{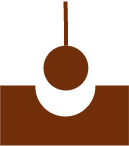}} 
        represents a module component, and \raisebox{-0.2em}{\includegraphics[height=1.2em]{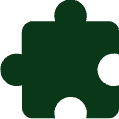}} represents a plugin.}
    \label{fig:framework}
\end{figure}


\subsection{Example for Illustrating Functionalities of the Modules and Plugins}
\label{sec:example}

We present an example to simulate a campus containing multiple agents, where the \module{Agent}, \module{Environment}, and \module{Action} modules are used to define agent behaviors, manage the shared world, and provide agent capabilities, respectively. The interaction follows a clear pattern: an \module{Agent} perceives the \module{Environment} and selects an \module{Action} to execute, which further modifies both the \texttt{Agent}'s internal state and the \module{Environment}. 
The core modules of \module{Controller} \module{System} coordinate the entire simulation process.
We now examine each module for the campus-simulation scenario in detail.

The \module{Agent} module drives the behavior of each agent through an internal suite of plugins. Specifically, an agent's behavior typically follow the following routine: \plugin{Perceive $\rightarrow$ Plan $\rightarrow$ Invoke $\rightarrow$ State  $\rightarrow$ Reflect}. Note that the routine is configurable, allowing users to adjust the logic and execution order of plugins, such as adding a contemplative agent that reflects before invoking. For example, consider a student, Alice, who has a Profile of ``fresh-year student majoring in Computer Science''. The \plugin{Perceive Plugin} receives a ``10 AM class reminder''. Based on this perception, the \plugin{Plan Plugin} generates a goal: ``go to the classroom''. The \plugin{Invoke Plugin} executes a ``move'' action, which updates her coordinates as she walks. If she receives a message from another agent such as Bob, the \plugin{Plan Plugin} might decide to reply based on her current ``social need'' state, and the \plugin{Invoke Plugin} would execute the \method{send\_message()} action. After these actions, the \plugin{State Plugin} updates her internal state, decreasing ``energy'' from ``high'' to ``tired''. At the end of the day, the \plugin{Reflect Plugin} reviews the day's events, summarizing conversations to form new memories like ``had a great chat with Bob today''.

The \module{Environment} module defines the physical and social space of the entire campus. It manages plugins like the \plugin{Space Plugin} , which defines the coordinates of buildings such as the library and dormitories. It also manages a \plugin{Relation Plugin} that dynamically handles the social network among agents. The power of this design is that users can completely change the environment by using different plugins, without modifying the code of agent. For example, users could upgrade the \plugin{Space Plugin} from 2D to 3D coordinates to enrich movement behaviors. Alternatively, an upgraded \plugin{Relation Plugin} could propagate the content of conversations to third parties with a certain probability, simulating a realistic ``gossip'' effect.

The \module{Action} module provides a collection of capabilities that agents can perform. In this way, an agent's abilities become modular, making them easy to register, discover, invoke, and extend. It manages plugins like the \plugin{Communication Plugin}, which defines the implementation of ``private chat'' and ``group chat'', allowing for realistic dialogue between students. It also provides agents with tools like a \plugin{Course Schedule Query Plugin} and a \plugin{Library Search Plugin} to retrieve information. It can even integrate external tools that follow the standard Model Context Protocol (MCP), such as a real-world weather query tool, to influence the actions of an agent, such as  ``carrying an umbrella when going outside in rainy days''.

The \module{Controller} module acts as the central coordinator. For example, if a student attempts to perform a ``talk'' action during an exam, the controller will intercept the request to comply with the examination discipline. Another critical feature of the \module{Controller} is the ability of rolling back the entire simulation to a specific moment in time. This allows for ``what-if'' analysis, such as studying the impact of a sudden event on the social network.

The \module{System} module orchestrates the simulation's foundational elements. The \plugin{Timer} drives the passage of time in the campus world, measured in ticks. For example, 48 ticks represent 24 hours on campus. The \plugin{Messager} manages inter-agent communication. The \plugin{Recorder} tracks various statistical indicators, such as students' movements and conversations, enabling analysis of phenomena like differences in social circles among students from different majors.

Crucially, new possibilities can be added by simply registering one or more plugins with the relevant modules.
To introduce an ``exam'' scenario, for example, users could register a new ``exam-taking'' plugin with the \module{Action} module. This registration immediately grants all agents this new capability, which their \plugin{Plan Plugin} can now discover and schedule.
Simultaneously, the \module{Controller} can be configured to orchestrate the event. It leverages the \module{System} module's central \plugin{Messager} to dispatch a global ``exam start'' event to eligible students. The agents then perceive this event, which triggers their cognitive loop.
Furthermore, the \module{Controller} enforces exam discipline by acting as the mandatory gatekeeper for all interactions. It intercepts and blocks forbidden actions, such as ``talk'', thereby enforcing the scenario's rules.
Thus, a complex, rule-based social phenomenon is introduced entirely by adding isolated plugins and configuration, without modifying the framework's core kernel.

The campus-simulation example comprehensively demonstrates the capabilities of Agent-Kernel, highlighting the primary advantage of its microkernel architecture: a stable module core system with customizable and extensible plugins.

\subsection{Comparison with Other MAS Frameworks}
\label{sec:comparison_with_other-ach}

\newcommand{\FSzero}{-}
\newcommand{\FSone}{\FiveStar}
\newcommand{\FStwo}{\FiveStar\FiveStar}
\newcommand{\FStree}{\FiveStar\FiveStar\FiveStar}
\newcommand{\FS}[1]{%
  \ifcase#1\relax
    \FSzero
  \or\FSone
  \or\FStwo
  \or\FStree
  \else
    \FiveStar\FiveStar\FiveStar
  \fi
}

We compare Agent-Kernel with representative MAS frameworks—spanning the pipeline and layered architectures discussed in Section~\ref{sec:introduction}—based on a comprehensive survey of their architectural support for realistic social simulation across four critical dimensions: adaptability, configurability, reliability, and reusability. Table~\ref{tab:mas_frameworks} presents the quantitative results of this comparison. Note that the ratings are results from our assessment through extensive hands-on experience with these frameworks, interpreting their design philosophies and documented features in the context of social simulation requirements.

\begin{table}[t]
\centering
\caption{%
  Comparative capabilities of MAS Frameworks for Social Simulation across four dimensions. 
  The rating ranges from 1 to 3 stars, with 3 stars indicating the highest capability, and ``-'' indicating no capability.
}
\label{tab:mas_frameworks}
\footnotesize
\begin{tabular}{
@{}>{\raggedright\arraybackslash}p{3.5cm}      
>{\centering\arraybackslash}p{2cm}            
>{\centering\arraybackslash}p{2cm}            
>{\centering\arraybackslash}p{2.2cm}          
>{\centering\arraybackslash}p{2cm}            
>{\centering\arraybackslash}p{2cm}@{}         
}
\toprule
\textbf{Framework} &
\textbf{Architecture} &
\textbf{Adaptability} &
\textbf{Configurability} &
\textbf{Reliability} &
\textbf{Reusability} \\
\midrule
\textbf{Agent-Kernel (Ours)} & \textbf{microkernel} & \textbf{\FS{3}} & \textbf{\FS{3}} & \textbf{\FS{3}} & \textbf{\FS{3}} \\
AgentSociety~\cite{agentsociety2025} & layered & \FS{0} & \FS{2} & \FS{1} & \FS{2} \\
AgentScope~\cite{agentscope2025} & pipeline & \FS{0} & \FS{2} & \FS{1} & \FS{1} \\
CAMEL~\cite{camel2023} & layered & \FS{0} & \FS{2} & \FS{1} & \FS{1} \\
CrewAI~\cite{crewai} & pipeline & \FS{0} & \FS{1} & \FS{2} & \FS{1} \\
AutoGen~\cite{autogen2023} & pipeline & \FS{0} & \FS{1} & \FS{1} & \FS{1} \\
Concordia~\cite{vezhnevets2023generative} & layered & \FS{0} & \FS{2} & \FS{2} & \FS{2} \\




BabyAGI~\cite{babyagi} & pipeline & \FS{0} & \FS{1} & \FS{1} & \FS{1} \\
DeepResearchAgent~\cite{deepresearchagent} & pipeline & \FS{0} & \FS{1} & \FS{1} & \FS{1} \\
BotSharp~\cite{botsharp} & pipeline & \FS{0} & \FS{1} & \FS{1} & \FS{1} \\
LangGraph~\cite{langgraph} & pipeline & \FS{0} & \FS{2} & \FS{1} & \FS{2} \\
OpenManus~\cite{openmanus} & pipeline & \FS{0} & \FS{1} & \FS{1} & \FS{1} \\
MetaGPT~\cite{metagpt2023} & pipeline & \FS{0} & \FS{2} & \FS{2} & \FS{2} \\
PraisonAI~\cite{praisonai} & pipeline & \FS{0} & \FS{1} & \FS{1} & \FS{1} \\

BeeAI~\cite{beeai} & pipeline & \FS{0} & \FS{1} & \FS{2} & \FS{1} \\

Agents SDK~\cite{agentssdk} & pipeline & \FS{0} & \FS{1} & \FS{2} & \FS{1} \\
MindSearch~\cite{mindsearch2024} & pipeline & \FS{0} & \FS{1} & \FS{1} & \FS{1} \\
DeerFlow~\cite{deerflow} & pipeline & \FS{0} & \FS{2} & \FS{1} & \FS{2} \\

AgentVerse~\cite{agentverse2023} & pipeline & \FS{0} & \FS{2} & \FS{1} & \FS{1} \\

Claude-Flow~\cite{claudeflow} & pipeline & \FS{0} & \FS{2} & \FS{2} & \FS{2} \\
OxyGent~\cite{oxygent} & pipeline & \FS{0} & \FS{1} & \FS{2} & \FS{1} \\
Agno~\cite{agno} & pipeline & \FS{0} & \FS{2} & \FS{2} & \FS{2} \\
TradingAgents~\cite{tradingagents2024} & pipeline & \FS{0} & \FS{1} & \FS{2} & \FS{1} \\
ROMA~\cite{roma2020} & layered & \FS{0} & \FS{1} & \FS{1} & \FS{1} \\

\bottomrule
\end{tabular}
\normalsize
\end{table}

\textbf{Adaptability.} In this context, 3 stars denote full runtime population dynamics—the ability to add or remove agents during execution—whereas a ``-'' indicates a static population that cannot be altered after initialization. As shown in Table~\ref{tab:mas_frameworks}, Agent-Kernel is, to our knowledge, the first framework to achieve this via Agent-Environment-Action decoupling (Section 3.2), enabling true simulations of evolving societies with birth, death, and migration. In contrast, typical pipeline frameworks such as MetaGPT~\cite{metagpt2023} and CrewAI~\cite{crewai} rely on predefined Standard Operating Procedures (SOPs)  or static workflow definitions to orchestrate agent interactions.

\textbf{Configurability.} In this context, 3 stars denote the capability for runtime modification of parameters and societal configurations; 2 stars indicate global pre-run configuration tailored for social simulations; and 1 star refers to hard-coded settings or limited pre-run configuration at the agent level. Many agent-centric frameworks require modifications across numerous agent definitions to adjust global parameters. Agent-Kernel's modular architecture (Sections 3.2–3.3) enables rapid reconfiguration through declarative plugin specifications, while the Controller (Section 3.5) provides runtime interventions for efficient hypothesis testing. Conversely, frameworks like AutoGen~\cite{autogen2023} and CrewAI~\cite{crewai} often depend heavily on hard-coded configurations or static files (e.g., YAML), limiting flexibility for real-time adjustments. While some, like LangGraph~\cite{langgraph} and BotSharp~\cite{botsharp}, offer runtime intervention or visual management tools, they primarily focus on optimizing task execution flows rather than adjusting global social simulation parameters.

\textbf{Reliability.} We assign 3 stars for deterministic execution validated by strict logic (e.g., Timer/Controller mechanisms); 2 stars for safety mechanisms such  prompt auditing or configuration security checks; and 1 star for systems relying on asynchronous, non-deterministic interactions or merely basic error handling. Agent-Kernel ensures deterministic, reproducible results through the global Timer and Controller validation mechanisms (Sections 3.5–3.6), preventing the non-deterministic outcomes common in asynchronous agent-centric frameworks. Many existing frameworks, including AgentScope~\cite{agentscope2025} and CAMEL~\cite{camel2023}, primarily rely on LLM-based probabilistic outputs and asynchronous communication, which can lead to issues like infinite loops or role-flipping without strict deterministic controls. Although some frameworks like MetaGPT~\cite{metagpt2023} and Agno~\cite{agno} introduce mechanisms for self-correction or type safety, they often lack the rigorous, simulation-wide deterministic guarantees provided by a centralized kernel.

\textbf{Reusability.} Regarding reusability, we adopt a cumulative scoring criterion in which one star is awarded for each satisfied property: support for basic code-level reuse of agent classes or small functional units, availability of modular components tailored to social simulation contexts, and a ``hub'' ecosystem that enables community-wide sharing of simulation models. Agent-Kernel's plugin-based microkernel architecture (Section 3.3) and Database-per-Plugin strategy (Section 3.4) treat components as self-contained, interchangeable units that can be seamlessly reused across different simulation scenarios, enabling developers to build libraries of validated components. In comparison, frameworks such as PraisonAI~\cite{praisonai} and OxyGent~\cite{oxygent} typically offer reusability only at the level of code classes or functional tools, lacking a broader module-sharing ecosystem. While platforms like Claude-Flow~\cite{claudeflow} and MetaGPT~\cite{metagpt2023} have begun to establish cloud-based marketplaces or agent stores, their reuse mechanisms are often tied to specific application logic rather than generalizable social simulation modules.

It is worth noting that the landscape of MAS frameworks is vast and constantly evolving, making any comparison incomplete. Our analysis focuses on extensible frameworks; therefore, closed-source projects (e.g., SocioVerse) and end-user visualization applications or specific simulations (e.g., Dify~\cite{dify2024}, Coze~\cite{coze2024}, or the Stanford Smallville simulation~\cite{park2023generative}) are outside the scope of this comparison. Extant frameworks often have different target audiences, serving either enterprise or research communities. They are designed for specific domains (e.g., TradingAgents~\cite{tradingagents2024}) or are part of a larger proprietary solution (e.g., Agno~\cite{agno}). Furthermore, frameworks have derivatives or wrappers, or inherit from multiple predecessors, and some may use different underlying models. Therefore, assessing a framework's capability for social simulation inevitably involves a degree of subjective interpretation shaped by its architectural philosophy. For example, Concordia~\cite{vezhnevets2023generative} is classified as layered because its ECS design centralizes behavior within system modules rather than exposing it through a microkernel-style plugin interface. Nevertheless, the systematic advantages of Agent-Kernel's society-centric, modular microkernel architecture remain clear across all four dimensions.


\section{Modular Core System of Agent-Kernel}
\label{sec:agent_kernel}

Agent-Kernel overcomes the shortcomings of existing frameworks in adaptability, configurability, reliability, and reusability through a society-centric modular microkernel architecture, which consists of a stable, modular core system and a series of application-specific plugins. In this section, we detail the design of the modular core system, which serves as the foundational scaffolding for all simulations and comprises five essential modules: \module{Agent}, \module{Environment}, \module{Action}, \module{Controller}, and \module{System}. To structure this analysis, We begin by articulating the essential requirements for constructing realistic and reliable social simulations, while highlighting the challenges imposed by the limitations of existing frameworks. Subsequently, we detail the design of Agent-Kernel’s core modules, explicitly demonstrating how our architecture effectively resolves these challenges.

\subsection{Challenges for Social Simulation}
\label{sec:challenges}
Realistic and adaptive social simulations present multifaceted challenges. As outlined in Section~\ref{sec:introduction}, these systems demand the simultaneous satisfaction of four critical requirements: adaptability, reusability, reliability, and configurability. We have compared existing systems along with these dimensions in Section~\ref{sec:comparison_with_other-ach}. The specific challenges are discussed below.

\textbf{Adaptability}. Societies are inherently dynamic, with shifting populations, evolving actions, and ever-changing environments. To faithfully capture this complexity, social simulations must be able to adapt to such changes. Achieving such adaptability requires that a MAS system support real-time updates to agents, actions, and environments. However, most existing MAS frameworks employ an agent-centric design, in which environmental information and action lists are tightly coupled to individual agent profiles. As a result, even minor updates often require modifying a large number of agents. For example, adding a new agent necessitates updating the social relationship lists of all existing agents that may interact with it. These structural rigidities make the realization of truly adaptive social simulations an enduring challenge.

\textbf{Reusability}. Societies exhibit tremendous diversity, with a wide range of structural and functional variations across different contexts. Building simulations for such varied scenarios is often costly, as it typically requires reconstructing foundational components for each new study, resulting in significant resource waste. To address this and enhance development efficiency, social simulations must support reusability, enabling researchers to assemble diverse scenarios by leveraging common functional modules. Yet, most existing frameworks fall short in this regard, forcing developers to repeatedly reimplement shared features. There is, therefore, a pressing need for infrastructure that prioritizes modularity to eliminate such redundant development efforts.

\textbf{Reliability}. Societies are inherently sensitive to mistakes—even a single misstep can trigger far-reaching changes in structure, dynamics, or collective trajectory. In LLM-based social simulations,  this vulnerability is amplified by the hallucination phenomenon inherent to Large Language Models, which frequently results in unpredictable errors. For instance, an agent may formulate a plan that involves an infeasible action or violates established social norms. If left unchecked, such errors often trigger cascading failures that cause the entire simulation to crash or descend into chaos, thereby rendering the results meaningless. Consequently, ensuring reliability is paramount. However, most existing frameworks lack a unified supervisor to rigorously govern all agent behaviors, resulting in fragile systems prone to instability. Without a robust mechanism to contain these errors, the reliability required for meaningful scientific inquiry cannot be guaranteed.

\textbf{Configurability}. Social analysis often requires adjusting experimental configurations and systematically observing the resulting changes to gain insights. For example, researchers may need to update an agent's profile to observe shifts in cognitive behavior, restructure social relationships to study emergent phenomena such as information diffusion, or redesign spatial layouts to analyze environmental impacts during execution. To support such dynamic hypothesis testing, the simulation system must offer configurability, allowing researchers to intervene and modify key parameters during runtime. However, most existing MAS frameworks only support simple interactions, such as sending surveys, and lack the ability to directly alter fundamental simulation states—like agent attributes or environmental structures—once the simulation has started. Achieving comprehensive configurability, particularly the ability to intervene during runtime, thus remains a significant hurdle for existing systems.

Realistic social simulations also require robust system-level infrastructure. Regarding \textbf{Data Management}, complex social simulations generate diverse, heterogeneous data types, ranging from social relationship networks and semantic memories to environmental spatial layouts. To store and access this information efficiently, the system must support heterogeneous persistence tailored to specific data characteristics. However, existing frameworks often rely on a centralized, shared database approach. Consequently, this forces a rigid "one-size-fits-all" model that creates system-wide fragility—a schema change in one functional module can trigger cascading errors that destabilize the entire system. \textbf{System coordination} mechanisms poses equally critical challenges. Large-scale simulations involve massive concurrent interactions that must remain logically consistent and traceable. To guarantee validity, the system requires strict alignment mechanisms for time synchronization, message passing, and data logging. However, existing frameworks often lack such unified coordination services. Consequently, asynchronous execution leads to fatal ``causal inversion" where faster agents act on ``future" information, while synchronous communication risks large-scale deadlocks, and the absence of centralized logging renders subsequent scientific analysis impossible due to fragmented data records.


\subsection{Agent-Environment-Action Decoupling}
To address the adaptability challenge rooted in the agent-centric design paradigm outlined in Section~\ref{sec:challenges}, we adopt a society-centric architecture that implements Agent-Environment-Action decoupling. This mechanism extracts the environment and actions from individual agents and treats them as society-level, shared abstractions managed by centralized modules. This principle strictly separates an agent's internal \textbf{cognitive} faculties from the external, \textbf{non-cognitive} world. Here, cognitive aspects refer to the subjective, internal world of each agent, including its fundamental profile, its dynamic state, and its core cognitive processes of perception, planning, invocation and reflection. In contrast, non-cognitive elements represent the objective, shared reality, including spatial layouts, social structures, and the definitions of physical behaviors. We illustrate this design using examples from our campus simulation (detailed in Section~\ref{sec:example}).

Under this decoupling, the \module{Agent} module encapsulates only the cognitive aspects. It maintains each agent's profile and dynamic state (e.g., energy, mood) while orchestrating its behavior through a structured cognitive loop. This loop, inspired by classic AI models like BDI (Belief-Desire-Intention) and recent advancements in LLM agents \cite{velleman1991intention,park2023generative}, follows a ``Perceive-Plan-Act-Reflect'' cycle that models rational decision-making: agents perceive their surroundings, formulate plans based on their internal state and goals, select an appropriate action, and reflect on outcomes to update their beliefs. For instance, a student agent like Alice perceives a ``10 AM class reminder'' (an external event). Her \plugin{Plan Plugin}, guided by her academic goals, generates a plan: ``go to the classroom''. 

The non-cognitive elements are managed by two centralized modules. The \module{Environment} module serves as the single source of truth for the objective world and manages plugins that define the shared context, such as a \plugin{Space Plugin} for spatial locations and a \plugin{Relation Plugin} for the dynamic social network. This centralization ensures that Alice does not maintain her own version of the map; instead, she perceives the single, authoritative map from the \module{Environment} module. For example, the \plugin{Space Plugin} can be updated to include a newly constructed library building, and all agents will immediately perceive this new location.

Similarly, the \module{Action} module acts as a central repository of capabilities that agents can execute. Rather than treating actions as innate abilities hardcoded into individual agents, this design treats them as shared, discoverable capabilities available to the entire society. It contains foundational action plugins including the \plugin{Communication Plugin} for inter-agent interaction, the \plugin{Tools Plugin} for capability extension, and the \plugin{Other-Actions Plugin} for general virtual behaviors. When Alice's planning process decides to ``talk'', it invokes the ``communication'' action defined in this module, making capabilities both modular and discoverable across all agents.

By centralizing environmental and behavioral definitions, the architecture fundamentally supports dynamic population changes such as birth, death, and migration. Adding a new agent no longer necessitates ubiquitous updates to the internal states or relationship lists of existing agents. Instead, it requires only a simple registration within the \module{Environment}'s plugins (e.g., the \plugin{Relation Plugin}). Once registered, the new agent immediately inherits the ability to perceive the shared world and utilize all available actions. This mechanism ensures that the simulation remains stable and consistent even as the population scales or fluctuates.

\subsection{Plugins Management}
To systematically address the reusability challenges identified in Section~\ref{sec:challenges}, Agent-Kernel adopts a strategy of encapsulating scenario-specific logic into plugins, effectively separating stable core functions from variable simulation needs. However, this approach introduces potential fragmentation; as users develop plugins for diverse scenarios, these components often exhibit inconsistent interfaces and varying forms. This lack of a unified standard significantly complicates the management of the core system. To resolve this, Agent-Kernel implements a robust Plugins Management architecture. We embed specialized management components into the core modules—specifically \module{Agent}, \module{Action}, and \module{Environment}—that serve as standardized sockets for connecting external plugins. Crucially, by defining mandatory abstract interfaces (e.g., \method{init()}, \method{execute()}), we ensure that these components can uniformly recognize and orchestrate any plugin conforming to the protocol. This strict standardization effectively decouples the system's stable control logic from complex, variable business logic.

Furthermore, this standardization transforms plugins of the same type into interchangeable units, directly enabling the dynamic runtime replacement capability. Since plugins share an identical interface signature, users can seamlessly swap them via the component's management interface without interrupting the ongoing simulation. This capability allows for instant adjustments to agent behaviors—such as replacing a rule-based \plugin{Plan Plugin} with an LLM-based one—thereby fulfilling the requirement for flexible, high-fidelity social simulations.

\subsection{Database-per-Plugin}
To address the data management challenges identified in Section ~\ref{sec:challenges}, Agent-Kernel adopts a decentralized Database-per-Plugin strategy. Inspired by the "Database-per-Service" pattern from microservice architecture, this core principle delegates all data storage and management responsibilities directly to the plugin, acting as the fundamental unit of data ownership. 

This approach enables heterogeneous persistence, allowing each plugin to select the most suitable storage solution based on its specific data characteristics. For instance, a \plugin{Relation Plugin} managing complex social networks can utilize a graph database for efficient relationship querying, while a \plugin{Reflect Plugin} handling long-term memory can employ a vector database for high-speed semantic search. By binding data to its respective plugin, we establish a clear domain of responsibility that achieves fault isolation: data anomalies are confined to the specific plugin, preventing cascading failures and significantly reducing debugging complexity. Furthermore, this isolation ensures that schema changes remain strictly local, preventing the chain reactions common in centralized storage designs and thereby safeguarding the system's overall stability.

\subsection{Controller Module}
To address the reliability risks and limited runtime configurability identified in Section ~\ref{sec:challenges}, Agent-Kernel introduces a lightweight \module{Controller} module. This module serves as a central hub to prevent the chaos caused by invalid behaviors and enable the deep real-time interventions missing in existing frameworks.

The \module{Controller} is stateless and primarily handles two functions: validation and runtime intervention. Every action request from any agent must pass through the \module{Controller}, which validates it against predefined rules before forwarding it to the appropriate module for execution. For example, when an agent attempts to move to a new location, the \module{Controller} first verifies whether the target location exists and is accessible before routing the request to the \plugin{Space Plugin}. Furthermore, the \module{Controller} provides a unified interface for runtime interventions, allowing researchers to inject commands, modify agent states, or adjust environmental parameters during simulation—such as issuing a warning to a specific agent or altering social ties to study emergent phenomena.

This centralized coordination design provides significant advantages. First, it enhances reliability by intercepting invalid or logically inconsistent actions at their source, preventing chaotic system states and ensuring the stability of simulation results. Second, it improves configurability by enabling dynamic hypothesis testing and real-time scenario adjustments without requiring system restarts. Importantly, because the \module{Controller} is stateless and focuses solely on validation and forwarding, it can be horizontally scaled to avoid becoming a performance bottleneck.

\subsection{System Module}
To address the critical system coordination challenges emphasized in Section 3.1, Agent-Kernel provides a centralized \module{System} module. This module integrates three essential services—\plugin{Timer}, \plugin{Messager}, and \plugin{Recorder}—to systematically resolve the challenges of causal inversion, communication deadlocks, and fragmented data logging.

First, a global \plugin{Timer} acts as the central clock, ensuring that all agents operate on the same time benchmark. This strict synchronization eliminates causal paradoxes where faster-running agents might act on ``future" information, thereby guaranteeing deterministic simulations. Second, the \plugin{Messager} provides an asynchronous message-passing mechanism. Rather than requiring agents to communicate directly—which risks deadlocks due to synchronous waits—messages are routed through the \plugin{Messager}, which handles delivery asynchronously. This design decouples senders from receivers, ensuring a robust communication flow that stabilizes the system under high load while effectively preventing potential deadlocks. Third, the \plugin{Recorder} offers a unified logging interface that captures all simulation events in a standardized format. By resolving the issue of chaotic, fragmented data, it creates a complete, traceable audit trail that is crucial for debugging, result validation, and ensuring the reproducibility of scientific findings.


\section{Plugins of Agent-Kernel}
\label{sec:plugin}

Plugins are the fundamental units for implementing the framework's functionality. Developers define the simulated world with specific behavioral logic and rules by writing or selecting different plugins. This section introduces the plugin designs for the framework's three core modules: \module{Agent}, \module{Environment}, and \module{Action}.

\subsection{Plugins for Agent Module}
The plugins within the \module{Agent} module construct a complete individual connitive process, including the \plugin{Profile Plugin}, \plugin{State Plugin}, \plugin{Perceive Plugin}, \plugin{Plan Plugin}, \plugin{Invoke Plugin}, and \plugin{Reflect Plugin}. An agent first perceives the world through the \plugin{Perceive Plugin}. Subsequently, the \plugin{Plan Plugin} generates a plan by synthesizing perceptual information from the \plugin{Perceive Plugin}, identity information from the \plugin{Profile Plugin}, and state information from the \plugin{State Plugin}. The \plugin{Invoke Plugin} then executes concrete actions based on this plan and triggers the \plugin{State Plugin} to modify the agent's state. Finally, the \plugin{Reflect Plugin} generates reflections that guide future actions. This design aligns with the classic paradigm for LLM-based agents and provides a solid foundation for reliable simulations from individual behaviors to collective phenomena.

\plugin{Profile Plugin}. The \plugin{Profile Plugin} manages an agent's long-term, static identity information, such as its name and gender, which form the foundation for its behavioral consistency. The plugin loads the agent's profile and provides interfaces to access and modify this data.

\plugin{State Plugin}. In contrast to static attributes, the \plugin{State Plugin} governs an agent's dynamic internal states, like emotion or health points. It loads the agent's real-time status and offers interfaces that other plugins use to update the state. The plugin can also contain internal logic for state change, such as an energy level that periodically decreases over time.

\plugin{Perceive Plugin}. The \plugin{Perceive Plugin} acts as the agent's sensory system. It collects the latest environmental information and messages. It then organizes these raw inputs into a standardized data structure, forming a coherent perception for use by subsequent cognitive stages.

\plugin{Plan Plugin}. The \plugin{Plan Plugin} synthesizes all available information (e.g., perceptual data, identity details, and current state) to generate a concrete action plan by invoking a large language model. This plan provides guidance for the agent's subsequent actions.

\plugin{Invoke Plugin}. The \plugin{Invoke Plugin} is responsible for execution. It first parses the content of the plan, translating the text-based instructions into specific, executable actions. It then calls interfaces from the \module{Action} module to carry them out.

\plugin{Reflect Plugin}. The \plugin{Reflect Plugin} synthesizes and summarizes an agent's past experiences to generates high-level insights. This process endows the agent with the capacity for self-reflection and provides valuable context for future planning.

\subsection{Plugins for Environment Module}
The plugins within the \module{Environment} module establish a shared, simulated world for the agents. This world provides a unified foundation for their perception, interaction, and action. In the current version, the \module{Environment} module contains the \plugin{Space Plugin} and the \plugin{Relation Plugin}, which are the most frequently used plugins in social simulation scenarios. The \plugin{Space Plugin} constructs the physical environment, while the \plugin{Relation Plugin} builds the social relationship network. However, the framework allows developers to extend it with other plugin types to meet the demands of more complex simulation scenarios.

\plugin{Space Plugin}. This plugin governs the spatial structure of the simulated world, managing the states of both static objects and dynamic entities (i.e., agents). It maintains the positional data for all entities within the environment. Furthermore, it provides essential functionalities such as spatial queries (e.g., retrieving nearby entities) and state modifications (e.g., updating an agent's location). 

\plugin{Relation Plugin}. The \plugin{Relation Plugin} focuses specifically on the network of social relationships that agents require for communication. It manages the various types and attributes of these relationships and supports both their querying and dynamic updating.

\subsection{Plugins for Action Module}
The plugins within the \module{Action} module define non-cognitive behaviors related to agent interaction and state changes, including the \plugin{Communication Plugin}, the \plugin{Tools Plugin}, and the \plugin{Other-Actions Plugin}. The \plugin{Communication Plugin} manages communicative behaviors between agents, the \plugin{Tools Plugin} provides tools, and the \plugin{Other-Actions Plugin} contains virtual behaviors for simulation scenarios. These plugins provide a foundation for agent response and interaction in complex environments.

\plugin{Communication Plugin}. The \plugin{Communication Plugin} manages all behaviors related to information transfer among agents. It covers core functions such as message encapsulation, dispatch, storage, and query. The \plugin{Communication Plugin} centrally manages all actions related to inter-agent communication. This approach ensures reliable communication and provides a basis for studying how information propagates in a simulated society.

\plugin{Tools Plugin}. The \plugin{Tools Plugin} manages tools that agents can use to extend their capabilities, including two types of tools. One type is local function tools; for these, a developer only needs to write a stateless Python function and its corresponding Docstring. These functions then become tools that agents can discover and use. The other type is remote service tools which follow the standard Model Context Protocol (MCP). A corresponding remote service proxy automatically discovers and registers all such tools.

\plugin{Other-Actions Plugin}. The \plugin{Other-Actions Plugin} manages certain virtual behaviors that agents perform in the simulation, such as moving, eating, and sleeping. These behaviors do not belong to the previous two categories. They typically involve general actions that relate only to the agent's own state. This provides a foundation for simulating a complete individual lifecycle for the agent.

Additionally, methods in the \module{Action} module’s plugins use annotations to distinguish between agent-callable actions and system-only management interfaces. By clearly distinguishing the intended caller for each method, the annotations enable permission control for actions and significantly improve code readability and maintainability.

\section{Implementation of Agent-Kernel}
\label{sec:implementation}

To realize the core design goals of the Agent-Kernel framework, we made a series of deliberate software design decisions (as illustrated in Fig. \ref{fig:software-design}). These decisions are implemented through specific software patterns and technological choices, such as the Ray framework for distributed computing and \code{asyncio} for efficient, non-blocking operations. This section first examines the implementation of the modular core system, detailing its five foundational modules. Subsequently, it delves into the plugins implementation and the distributed mode, explaining how the framework supports flexible plugins injection and scales to large-scale simulations.  

\begin{figure}[htbp]
    \centering
    \includegraphics[width=1.0\linewidth]{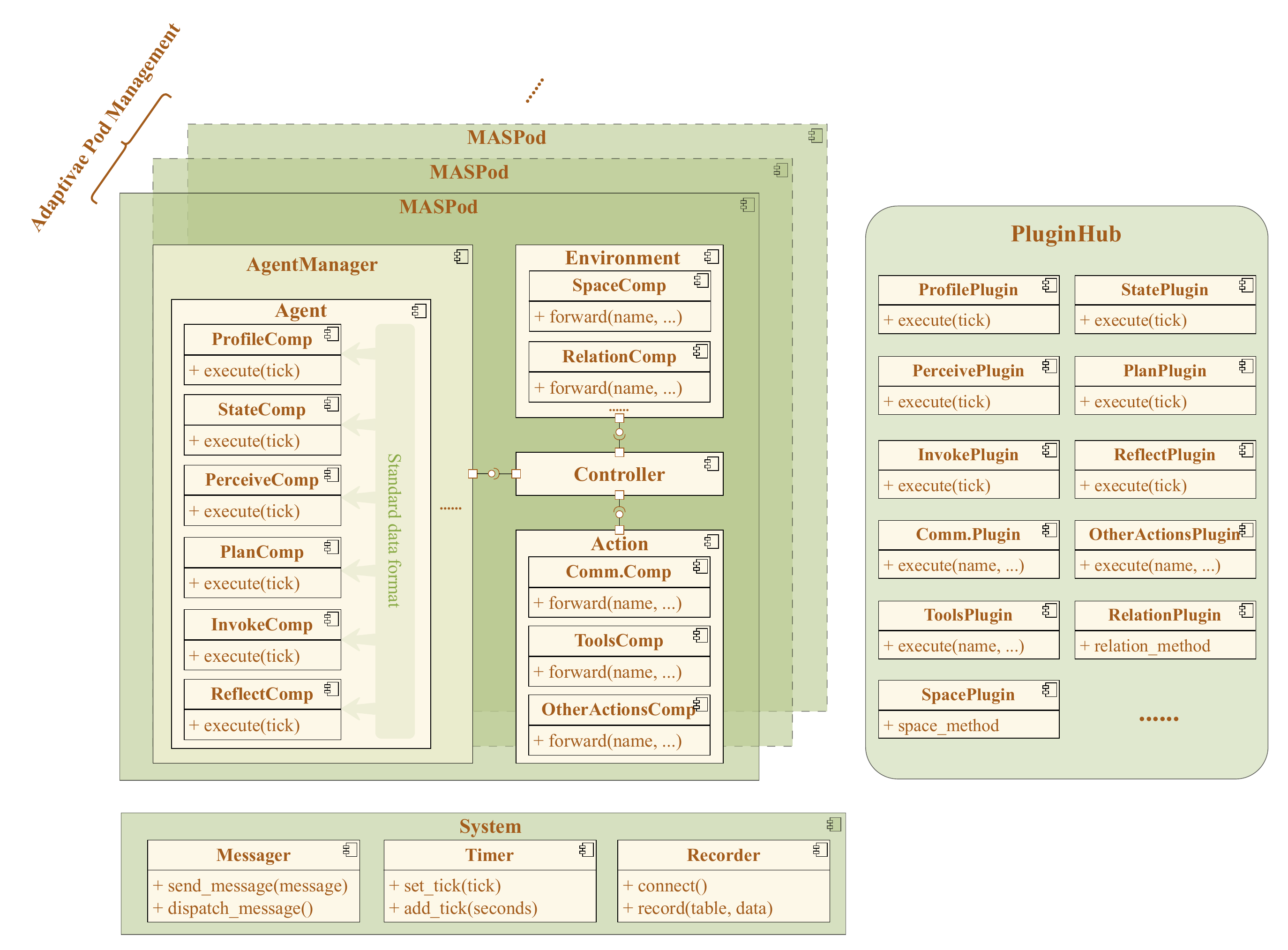}
    \caption{Software Design of the Agent-Kernel framework.}
    \label{fig:software-design}
\end{figure}

\subsection{Modular Core System Implementation}
The Agent-Kernel framework builds a modular core system with five modules: \module{Agent}, \module{Environment}, \module{Action}, \module{Controller}, and \module{System}. The implementation of this core system uses the Mediator pattern to keep modules loosely coupled and the Component-Plugin pattern to ensure high extendibility and reusability. These patterns operate directly at the code level to realize the framework's design goals. 

\textbf{Agent Module}. The \module{Agent} module is managed by an \class{AgentManager} class, which orchestrates the simulation tick for the agent population. At each tick, it iterates through all \class{Agent} instances and invokes their \method{run} method. The \class{Agent} class itself functions as a composite object, managing a collection of \class{AgentComponent} instances. The agent's behavior emerges from the sequential execution of these components (through \method{execute()}), each containing a hot-swappable \class{AgentPlugin}. Furthermore, each component caches data from its plugin as public properties using pre-defined data schemas (Pydantic models), making important data like perceptions or plans available to other components in the execution chain.

\textbf{Environment Module}. The \module{Environment} module follows the Facade pattern. The primary \class{Environment} class serves as a unified entry point to a collection of specialized \class{EnvironmentComponent} instances (e.g., \class{SpaceComponent}, \class{RelationComponent}). The \class{Environment} facade exposes a generic \method{run} method that acts as a dynamic dispatcher. This method receives a component name and a method name from the \module{Controller}, and subsequently routes the call to the appropriate internal component. This design allows the environment to be easily extended with new functionalities in future versions by simply adding a new component and its corresponding plugin, without altering the core framework.

\textbf{Action Module}. The \module{Action} module also follows the Facade pattern, similar to the \module{Environment} module. The \class{Action} class is a central registry for \class{ActionComponent} instances. Upon initialization, each component introspects its plugins for methods marked with the \annotation{AgentCall} annotation, which is used to enforce permission control by explicitly designating which methods are callable by agents. Based on these annotations, the component builds an internal routing table that maps a unique method name to the plugin that provides it. When the \module{Controller} invokes the \class{Action} facade's \method{run} method, this table is used to efficiently dispatch the call to the appropriate plugin.

\textbf{Controller Module}. The \module{Controller} module is the concrete implementation of the Mediator pattern, architected with an abstract \class{BaseController} defining basic interfaces and a \class{ControllerImpl} providing the concrete logic. The \module{Controller} holds direct references to the \class{AgentManager}, \class{Environment}, \class{Action}, and \class{System} objects, exposing a comprehensive \code{async} API as the exclusive channel for all inter-module communication. This asynchronous design is critical for performance, enabling the \module{Controller} to handle requests from numerous agents concurrently without blocking the main simulation loop.

\textbf{System Module}. The \module{System} module provides global, singleton-like services. It implements the core services \plugin{Timer}, \plugin{Messager}, and \plugin{Recorder} as standalone components, and these components become Ray actors in distributed mode. A non-actor \class{System} class acts as a service locator, and it holds handles to these components. The framework passes this lightweight object to each \module{Controller}, and this design gives each \module{Controller} a simple interface to global services. This approach also decouples all service users from the underlying implementation.

\subsection{Plugins Implementation}
The framework centers its extensibility on its plugin implementation, and it follows a consistent design pattern across the \module{Agent}, \module{Environment}, and \module{Action} modules. This approach lets developers integrate custom logic seamlessly when they follow two principles that define the system. The first principle uses type-specific abstract base classes to enforce a uniform interface. The second principle uses a dependency injection mechanism to manage database interactions through dedicated adapters.

Each plugin category enforces interface consistency through its own abstract base class (e.g., \class{PlanPlugin}), which defines a required \method{execute} method. This rule lets the parent component interact with any concrete plugin in a polymorphic way, facilitating code reuse and making plugins hot-swappable. For database connectivity, each plugin stays decoupled from the underlying database technology. Instead of connecting directly, each plugin receives \class{DatabaseAdapter} instances it needs during initialization and uses them to operate different databases. Each database adapter abstracts the database logic (e.g., for Redis or PostgreSQL) and provides a standard set of methods for data manipulation. This design centralizes data access logic within the adapters, and it keeps each plugin's business logic clean and database-agnostic.

\subsection{Distributed Mode Implementation}
To support large-scale simulations, the framework implements a distributed mode inspired by Kubernetes (K8s) principles of container orchestration. It builds this architecture on the Ray framework to distribute the computational load across multiple processes or machines. The design introduces two key concepts that define the distributed system. The first concept is the \module{MasPod} (Multi-agent System Pod), which acts as a deployment unit and encapsulates a group of agents together with their local \module{Environment}, \module{Action}, and \module{Controller} modules. The second concept is the \module{PodManager}, which manages the lifecycle, communication, and orchestration of all \module{MasPods}.

The \module{PodManager} follows the same design principle as the \module{Controller} module, comprising an abstract \class{BasePodManager} defining the interfaces and a concrete \class{PodManagerImpl} providing the logic. Both the \module{PodManager} and each \module{MasPod} are implemented as Ray Actors, enabling them to operate as independent, stateful units. The \module{PodManager} functions as the system’s central coordinator, which creates and manages all \module{MasPod} actors. It also serves as the exclusive message broker for inter-pod communication, and this rule keeps the control flow clean and centralized. Furthermore, the \module{PodManager} uses an adaptive strategy to balance the load across pods. When it receives a request to add a new agent, it queries the current agent count from each active \module{MasPod} and delegates the instantiation task to the pod with the fewest agents. This centralized control over both deployment and communication ensures that agents are evenly distributed, preventing computational bottlenecks and maximizing system throughput.

\section{Instructions for Developing Artificial Society Program with Agent-Kernel}
This section provides developers with a clear and concise guide to quickly building custom multi-agent social simulations using the Agent-Kernel framework. The process consists of the following steps:

1) \textbf{Write plugins for the core modules}. Developers first write plugins for the \module{Agent}, \module{Environment}, and \module{Action} modules that match the target simulation scenario. In plugins for the \module{Agent} module, developers define the agent's complete behavioral logic. In contrast, in plugins for the \module{Environment} and \module{Action} modules, they build a shared environment and provide executable actions or tools respectively.

2) \textbf{(Optional) Extend advanced functionality}. Developers can perform advanced extensions for complex scenarios that require global control or cross-module coordination. They can add new control methods, such as custom event broadcasting or access control, by inheriting from the base controller implementation class. In a distributed mode, they can also add new cluster coordination methods by inheriting from the \module{PodManager} implementation class. 

3) \textbf{Prepare initial data}. Developers prepare initial data before the simulation starts. They can use the Procedural Content Generation (PCG) tool from the framework's toolkit to automate this process. The process is simple: developers first define the content and distribution of data in a configuration file. Then, by running the PCG script, they can batch-generate simulated data files for agent identities, initial states, relationship networks, and spatial positions. Please note that PCG does not generate position information for static objects on the map; this file must be created by the developer.

4) \textbf{Deploy, configure, and monitor the simulation visually}. Finally, developers use the integrated web interface, the Society-Panel, to handle deployment, configuration, and monitoring. In this panel, They can upload code packages and generated simulated data files. They can then easily configure all modules using a visual form. After completing the configuration, they start the simulation with one click and monitor the runtime state in real time. The panel also allows them to send commands to the running simulation, e.g., dispatch a message to a specific agent or query its state.

5) \textbf{(Optional) Discover and reuse open-source projects on SocietyHub}. SocietyHub is a open-source community for multi-agent social simulation that promotes code reuse and collaboration. On this platform, developers can browse, filter, and discover various simulation projects contributed by the community. After reviewing a project's description and dependency information, they can download complete simulation packages. Then, they can deploy and use it in their local Society-Panel for learning or secondary development.

For a complete development guide with a detailed development workflow, configuration instructions, and some code examples, please refer to our GitHub repository: \href{https://github.com/ZJU-LLMs/Agent-Kernel}{https://github.com/ZJU-LLMs/Agent-Kernel}.

\section{Demonstrations}

We present two demonstrations to evaluate the key features of Agent-Kernel. The first follows the Universe 25 experiment and focuses on the framework's adaptability, while the second, the ZJU Campus Life simulation, highlights its capacity to support large-scale populations. 

To support the high-throughput inference demands of these simulations, we utilize a distributed pool of GPU resources. Specifically, the total computational power collectively serving both applications comprises: 8 NVIDIA A100 GPUs (80GB VRAM) hosting \code{Qwen/Qwen3-\allowbreak Next-\allowbreak 80B-\allowbreak A3B-\allowbreak Instruct}, 4 NVIDIA L40 GPUs (48GB VRAM) hosting \code{Qwen/Qwen3-\allowbreak 30B-\allowbreak A3B-\allowbreak Instruct-\allowbreak 2507}, and 4 NVIDIA RTX 4090 GPUs (24GB VRAM) hosting \code{Qwen/Qwen3-\allowbreak 30B-\allowbreak A3B-\allowbreak Instruct-\allowbreak 2507}.

\subsection{Simulation of Universe 25 Experiment}

In this section, we simulate the classic Universe 25 experiment to validate the framework's adaptability to varying population scales as well as its capability for modeling emergent and evolving behaviors.

\subsubsection{Simulation Scenario}
The Universe 25 experiment, conducted by John B. Calhoun in the 1960s–1970s, is a landmark study of mouse social behavior under high-density conditions. In a predator-free, resource-abundant enclosure, an initially small population expanded rapidly. As spatial density rose, behavioral pathologies emerged: parental neglect increased, male aggression intensified, and some mice withdrew from social and reproductive roles. These disruptions precipitated a broader breakdown of social organization, declining fertility, and eventual population collapse (Calhoun, 1973\cite{calhoun1973}).

Following the Universe 25 experiment, we leverage the framework's capabilities to conduct a conceptual simulation. Our design supports dynamic population sizing (expansion or contraction) and allows agents to interact across multiple generations. This structure enables the direct observation of how population-level behavioral patterns emerge, stabilize, or degrade, offering a controlled analogue to the complex social collapse observed in the original study.




\textbf{Environment Design.} Following the structural logic of the original Universe 25 experiment, we model the environment as an integrated spatial system. This system comprises key living components: food hoppers, water bottles, and cells (each housing nests). For computational efficiency, we compress the four-tiered nest structure into a single layer within each cell. This entire spatial configuration, along with the agents' social relationship networks, is managed by the \module{Environment} module within the Agent-Kernel framework.

\textbf{Agent Design.} We model each agent as a mouse (Mus musculus), consistent with the subjects of the Universe 25 experiment. Its attributes comprise two facets: an immutable profile and a mutable state. The profile anchors innate characteristics (e.g., gender, birth tick) as a static identity, whereas the state serves as a multi-dimensional mutable vector. To support complex behavioral simulation, we structure the state into three aspects:
\begin{itemize}
    \item \textbf{Physio-Social Basis:} Maintains physiological homeostasis, covering hunger, thirst, health, and energy, while also defining the agent's social identity, including colony affiliation and social role.
    \item \textbf{Psychological Metrics:} Encodes core variables that describe affective states, motivational tendencies, and stress-related conditions. These metrics include valence, arousal, prosociality, aggression urge, maternal motivation, social competence, acute stress, chronic stress, and learned helplessness.
    \item \textbf{Biological Lifecycle:} Defines the agent's developmental progression, specifying thresholds for key life stages: agents transition to independence at 1 day old, reach sexual maturity and social role assignment around 4 days old, enter senescence at 15 days old, and reach natural mortality around 20 days old. Concurrently, the estrous cycle is defined with a 2.5-day periodicity to align reproductive rhythms with the overall timeline.
\end{itemize}

\textbf{Action Design.} We design a total of 20 behaviors, categorized into two domains based on their interaction logic. The 8 social behaviors are managed by the \plugin{Communicate Plugin}. Conversely, the 12 individual behaviors are managed by the \plugin{Other Plugin}.

We configure a customized execution cycle for the agent as \plugin{Perceive $\rightarrow$ State $\rightarrow$ Plan $\rightarrow$ Invoke $\rightarrow$ State $\rightarrow$ Reflect}. The \plugin{Perceive Plugin}  captures nearby environmental context and interaction signals. The \plugin{State Plugin}  executes updates both following perception and invocation: the former simulates immediate reactions to environmental stimuli, while the latter resolves the comprehensive consequences of actions. Based on updated state, the \plugin{Plan Plugin}  generates three weighted candidate plans and supports adaptive re-planning, after which the \plugin{Invoke Plugin}  maps the selected plan to executable actions by calling upon the \module{Action} module. Finally, the \plugin{Reflect Plugin}  enforces survival checks based on agent attributes, while periodically generating reflections and updating social relationship networks.





\subsubsection{Results}
The simulation was initialized with an equal sex ratio of 4 pairs of mice (4 males, 4 females), all in a healthy state. Following an initial adaptation period, the population began a multi-generational expansion driven by births and deaths. The simulation ran for 1,729 ticks, equivalent to 73 simulated days. Over this duration, the complete life cycles of 447 individuals were recorded, resulting in a total of 126,265 logged behavioral events.

The following analysis first evaluates the relationship between execution time and the number of active agents, effectively demonstrating the framework's strong adaptability in handling dynamically changing computational loads. It then focuses on the population dynamics and behavioral patterns observed in the simulation, contrasting them with those of the Universe 25 experiment.

\textbf{Framework Adaptability.} Figure \ref{fig:runtime_dynamics_universe25} illustrates how execution time varies with the dynamically changing population. The bar chart for execution time and the line chart for population size exhibit a high degree of temporal synchronization. During the initial growth phase of the simulation, the rapid expansion of the agent population directly leads to a proportional increase in the computational load. Consequently, the execution time per tick rises sharply, attaining its maximum value concurrently with the population peak. Subsequently, throughout the population decline phase, the execution time decreases smoothly in parallel with the diminishing agent count.
The strong synchronization between execution time and agent population indicates that the framework efficiently and promptly responds to rapid changes in agent numbers, fully demonstrating its adaptability.
\begin{figure}[h]
    \centering
      \includegraphics[width=0.85\textwidth]{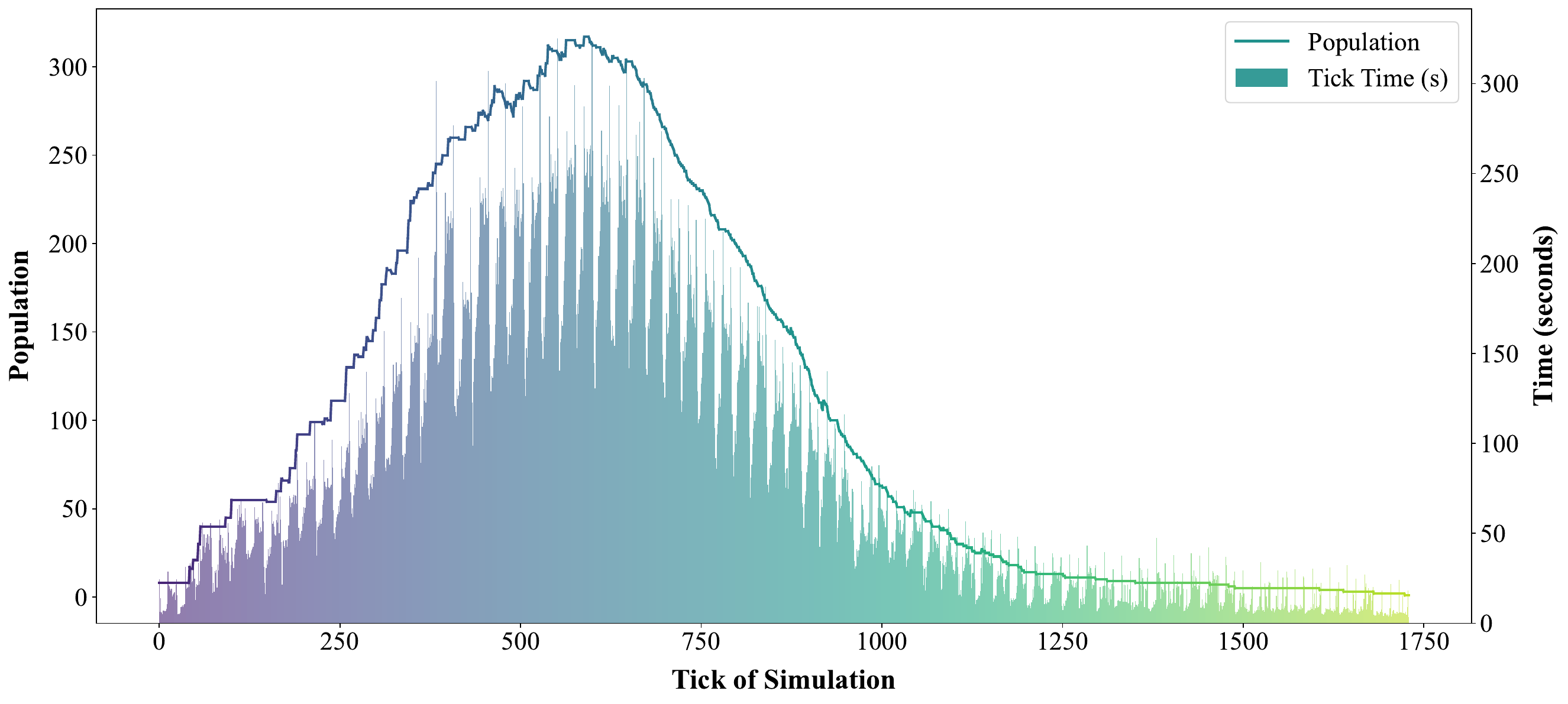}
    \caption{Runtime dynamics as a function of population size in the Universe 25 simulation.}
    \label{fig:runtime_dynamics_universe25}
\end{figure}

\textbf{Population, Behavior Patterns.} Population dynamics, including both trajectories and birth-death patterns, are illustrated in Figure~\ref{fig:univse25_population_curve}, while Figure~\ref{fig:universe25_action_category_proportions} presents the daily proportional changes in behavioral patterns. Consistent with Calhoun's definition, the simulation period begins with Day 0 as the adjustment phase (Phase A), followed by three subsequent developmental phases: Phase B (Days 1–14), Phase C (Days 15–23), and Phase D (Days 24–72).

\begin{figure}[htbp]
    \centering
    \includegraphics[width=0.9\linewidth]{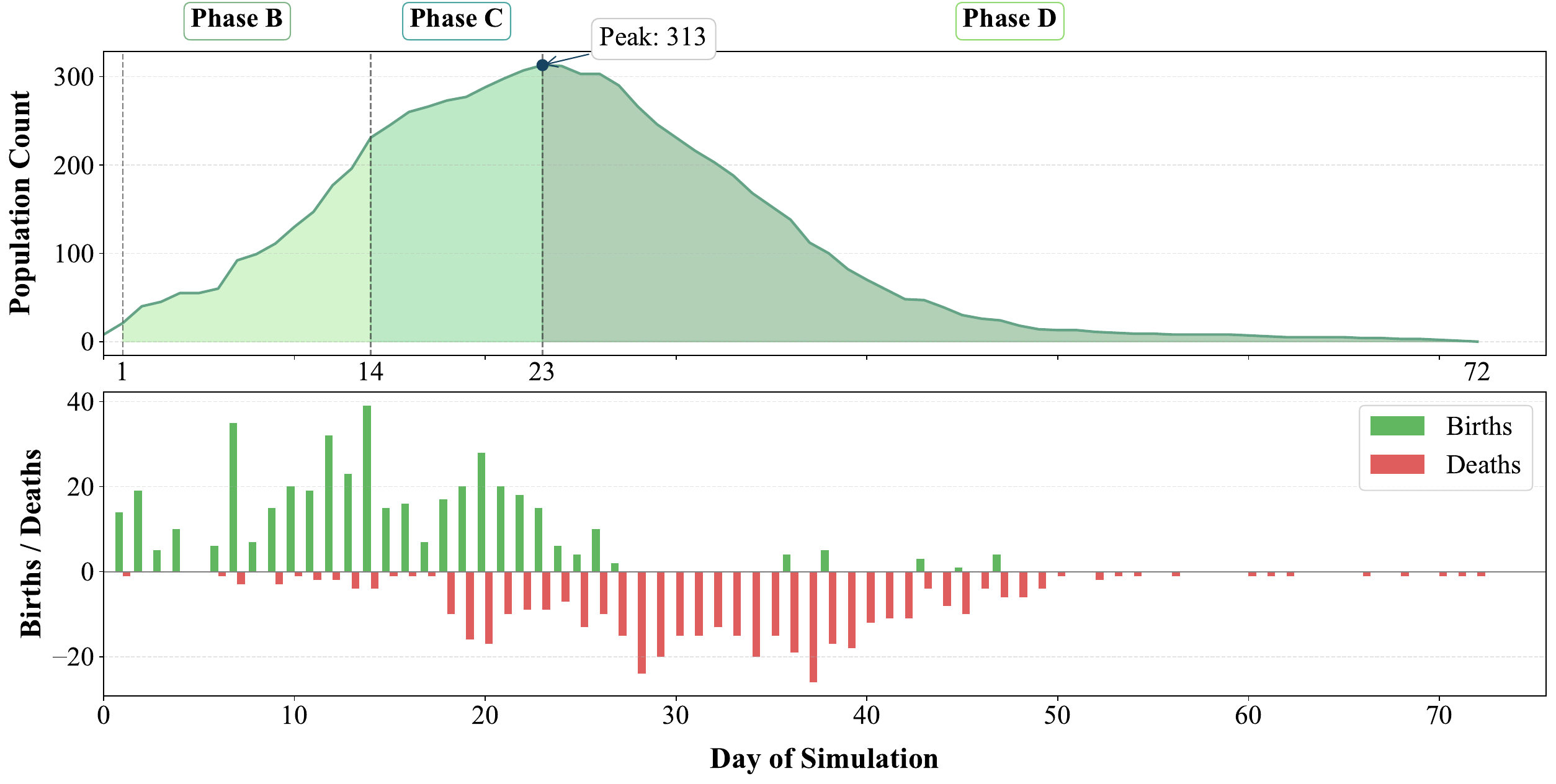}
    \caption{Population dynamics in the Universe 25 simulation.}
    \label{fig:univse25_population_curve}
\end{figure}

\begin{figure}[htbp]
    \centering
    \includegraphics[width=0.9\textwidth]{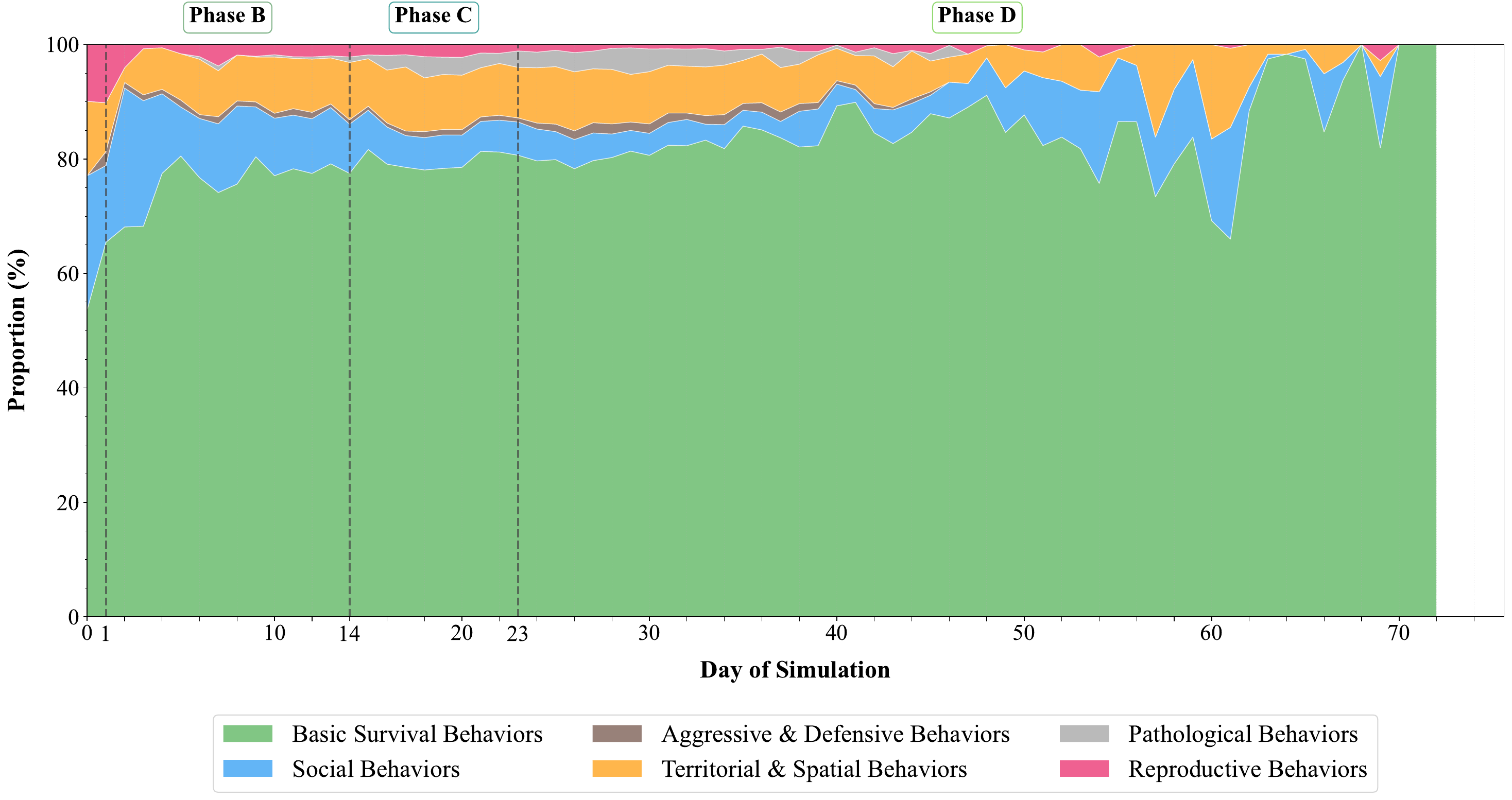}
    \caption{Proportion of daily behaviors in the Universe 25 Simulation.}
    \label{fig:universe25_action_category_proportions}
\end{figure}

During Days 1–14 (Phase B), the colony experiences rapid growth with 244 births against only 21 deaths, mirroring the exponential increase characteristic of the Explosive Growth Phase of the original Universe 25 experiment. The average behavioral patterns of this early phase showcase notable robustness. Social behaviors account for a strong average proportion of 12.44\%, territorial and spatial behaviors rank at 8.15\%, and reproductive behaviors are healthy at an average of 2.63\%. This combination collectively signifies the successful establishment of a stable, organized social structure under conditions of abundant resources and minimal social pressure, which strongly underpins the colony's demographic success.

Aligning with Calhoun's Stagnation Phase, the period from Day 15 to Day 23 (Phase C) shows population growth gradually leveling off, culminating in the population peak on Day 23. This phase is characterized by an average of approximately 17.33 births and 8.22 deaths daily, with the average daily net increase being 5 individuals lower than that recorded during Phase B.
The behavioral data in Figure~\ref{fig:universe25_action_category_proportions} reveals the social breakdown driving this stagnation: the proportion of social behaviors plummets by 53\%, and reproductive behaviors fall by 35\% compared to the previous stage.
Conversely, pathological behaviors, reflecting escalating psychological stress and functional disarray, surge to an average proportion of 2.49\%, while aggressive and defensive behaviors rank at 0.84\%. This evidence confirms that increasing social density and subsequent social pathology have become the new primary limiting factor on population growth.

After Day 23 (Phase D), the colony enters a sustained and irreversible decline, directly corresponding to the Decline Phase observed in the Universe 25 experiment. During Days 24–40 alone, mortality rises sharply to 274 deaths compared with only 31 births, and reproduction ceases entirely by Day 48, as shown in Figure~\ref{fig:univse25_population_curve}. This demographic collapse is caused by a profound shift in behavioral structure.
As illustrated in Figure~\ref{fig:universe25_action_category_proportions}, initial patterns involve widespread pathological and aggressive behaviors. However, over time, complex behaviors diminish as the patterns increasingly shift toward basic, solitary survival behaviors. This category of survival behavior increases to an average of 85.3\% and even reaches 100\% on several days, virtually eliminating all other behaviors essential for population maintenance.
Although the timeline exhibits fluctuations in complex behaviors, such as social and territorial behaviors briefly peaking near 20\% and 16\% respectively during Days 54–61 in Figure~\ref{fig:universe25_action_category_proportions}, these bursts reflect the last surviving mice's ultimately unsuccessful attempt to cope with accumulated pup-period stress. They soon become largely confined to the most fundamental physiological needs as well.
This extreme reversion mirrors the withdrawn state of Calhoun's "Beautiful Ones", confirming that the colony's eventual physical extinction results not from resource depletion but from the complete collapse of its complex social functioning.


\subsection{Simulation of ZJU Campus Life}

\subsubsection{Simulation Scenario}
The ZJU Campus Life simulation models a large-scale virtual society of 10,000 agents, each representing an individual person. We use this simulation to validate our framework's ability to model large-scale societies containing diverse agents, realistic environments, and rich interpersonal relationships.

\textbf{Environment Design.}
The \module{Environment} module is composed of two main components: the virtual campus map and the agent social network.
The virtual campus map is a two-dimensional replica of Zhejiang University's Zijingang Campus. It incorporates 41 representative locations that are grouped into five key functional areas: residences, sports, academics, administration, and landscapes.
The social network models the relationships among the individuals (agents) within the simulation. It comprises twelve distinct types of relationships, with each edge weighted according to its strength. On average, each agent maintains 27 connections, providing a realistic foundation for modeling social interactions.
In the simulation framework, the campus map is managed by the \plugin{Space Plugin}, while the social network is handled by the \plugin{Relation Plugin}.

\textbf{Agent Design.}
We synthesize 10,000 agents into four roles: 8000 students, 1000 faculties, 500 administrators, and 500 staff members. Agents share several common static attributes such as name, age, personality, and goals, while different roles of agents possess distinctive profiles reflecting their specific functions within the campus environment. In addition to these static profiles, each agent maintains five dynamic state attributes that are updated throughout the simulation, namely health, energy, happiness, stress, and social need.

\textbf{Action Design.}
The \module{Action} module organizes agent behaviors into eight categories: living, movement, academic, office, service, recreation, social, and sports. Each category contains several actions associated with specific agent roles, thereby governing their behavior in the simulation.
The majority of these actions are defined in the \plugin{Other Plugin}, which are performed individually and serve to influence the agents' internal states. A notable exception is the communication action defined in the \plugin{Communication Plugin}, which enables message exchange among agents. This specific action is crucial for simulating realistic social interactions and networking within the virtual environment.

Based on the Agent-Kernel framework, we model each agent's behavior within a single tick as a five-stage sequential pipeline: \plugin{Perceive $\rightarrow$ Plan $\rightarrow$ Invoke $\rightarrow$ State $\rightarrow$ Reflect}.
In each tick, an agent perceives the environment through the \plugin{Perceive Plugin}, formulates a plan using the \plugin{Plan Plugin}, selects an action from the \module{Action} module's plugins, executes it via the \plugin{Invoke Plugin}, updates its internal state via the \plugin{State Plugin}, and reflects on the outcome using the \plugin{Reflect Plugin}. To better emulate human-like memory processes, we implement a reflection mechanism in the \plugin{Reflect Plugin} that enables agents to condense conversations and daily activities into memories for future decisions.

\subsubsection{Results}
We distribute the 10,000 agents across 50 agent pods, with each pod managing 200 agents.
The simulation runs for 336 ticks, corresponding to 7 days in the virtual ZJU Campus. We first present a performance evaluation to assess the framework's ability to support large-scale simulations. We then provide behavioral validation through simulation-specific metrics and a case study, demonstrating the realism and validity of the simulated ZJU Campus environment.

\textbf{Time Cost.} 
The simulation completes 336 ticks in 29.9 hours, averaging 320 seconds per tick.
We break down a single tick into four phases: \emph{Agent Execution}, in which all agents concurrently run their full execution cycles; \emph{Message Dispatch}, responsible for serial delivery of inter-agent messages; \emph{Status Update}, where each all agents' states are updated concurrently; and \emph{Evaluation}, for computing simulation metrics.
Figure~\ref{fig:phase_and_token_timeline} presents the time distribution per tick across these phases, highlighting that the primary computational bottleneck occurs during \emph{Agent Execution}. This phase alone accounts for an average of 93.0\% of the total runtime. \emph{Message Dispatch} contributes 6.37\%, while \emph{Status Update} and \emph{Evaluation} incur only a minor fixed overhead per tick.

\textbf{LLM Token Usage.} 
The simulation consumes 1.26 billion tokens in total, averaging 3.75 million tokens per tick, demonstrating the framework's capacity to handle high-throughput LLM requests. Prompts dominate token usage, accounting for an average of 87.4\%, as they include large context information for planning, reflection, and communication, while completions constitute the remaining 12.6\%. Figure~\ref{fig:phase_and_token_timeline} shows that the token usage timeline closely aligns with the time consumption timeline, indicating that simulation performance is largely constrained by I/O operations for the LLM invocation requests.
Notably, both timelines exhibit clear periodic patterns: daily planning and reflection result in high token usage at the end of each day (e.g., ticks 46-47, 94-95), whereas inter-agent conversations occur occasionally during daytime, consuming comparatively fewer tokens.

\begin{figure}[!htbp]
    \centering
    \includegraphics[width=1.0\linewidth]{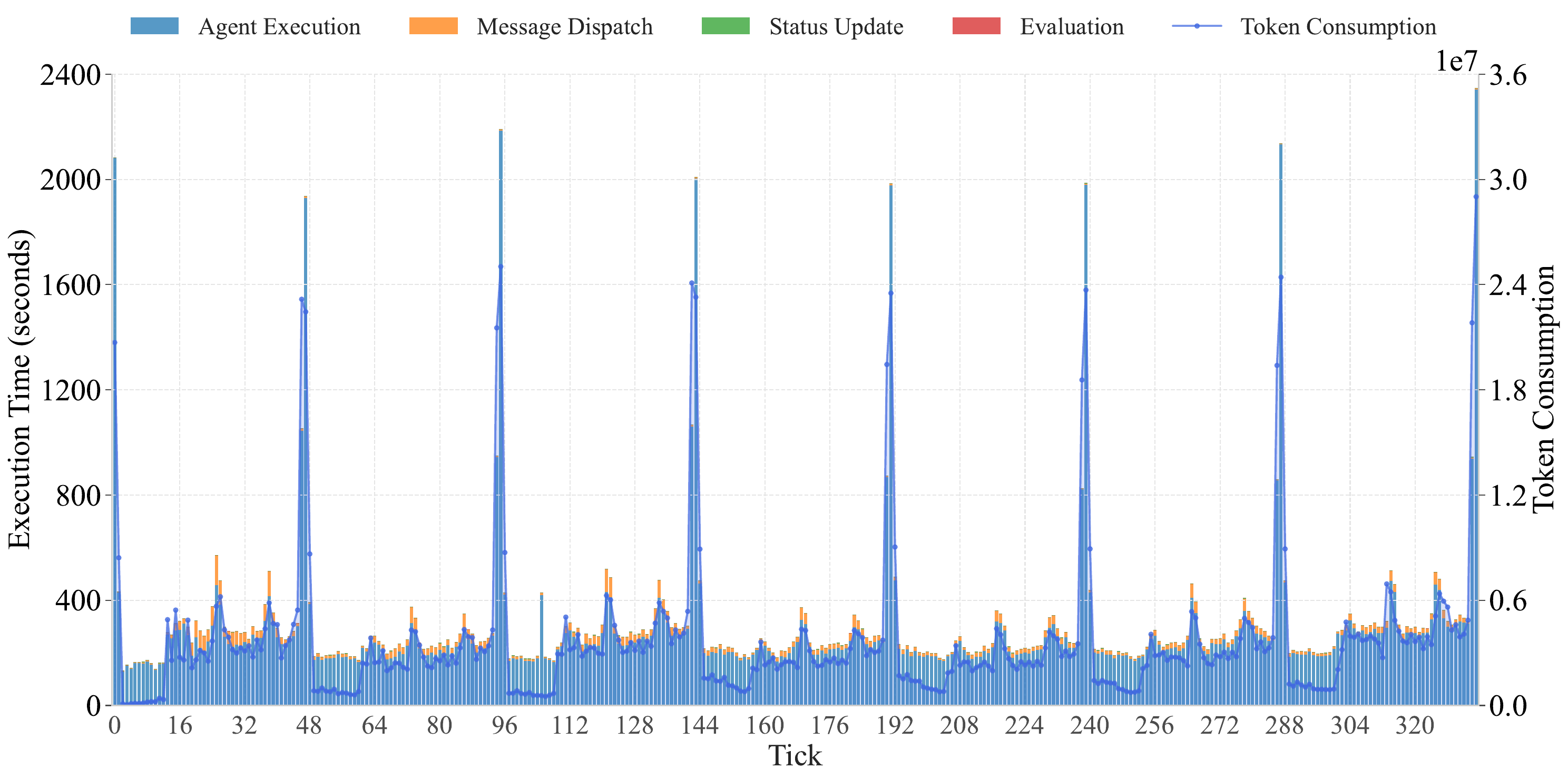}
    \caption{Token usage and phase-wise time breakdown per tick over the simulation timeline.}
    \label{fig:phase_and_token_timeline}
\end{figure}

\textbf{Memory Utilization.} 
Figure~\ref{fig:pod_workload_distribution} shows the workload distribution across all agent pods, with each pod consuming an average of 1.36 GB of RAM. The memory allocation is highly uniform, with a standard deviation of 67.71 MB, and a coefficient of variation of 4.86\%.
The low coefficient of variation indicates a highly uniform memory allocation, demonstrating that the framework's pod management strategy achieves strong load balancing.

\begin{figure}[!htbp]
    \centering
    \includegraphics[width=1.0\linewidth]{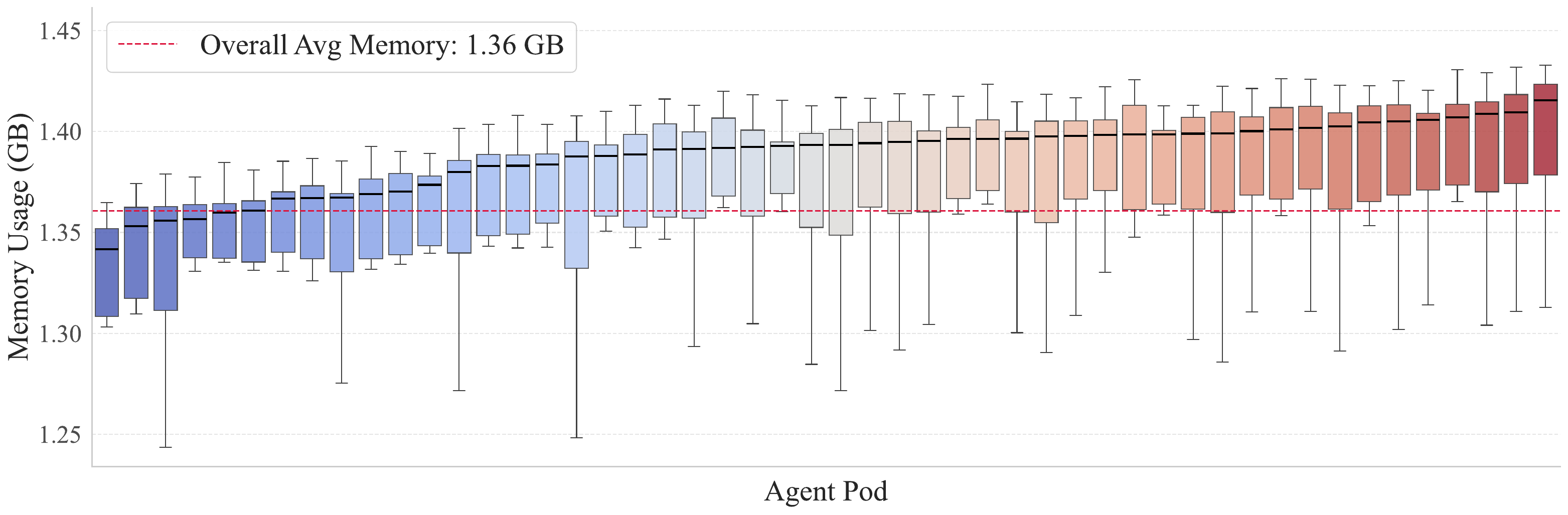}
    \caption{Workload distribution across 50 agent pods. Ordered by median memory consumption of each pod.}
    \label{fig:pod_workload_distribution}
\end{figure}

\textbf{Agent Behavior Analysis.}  
During the 7-day virtual ZJU Campus Life simulation, ten thousand agents performed 1.33 million actions and made 474 thousand visits to 41 locations.
Figure~\ref{fig:combined_student_plots} (left) shows that daily activities are dominated by living and academic actions, while movement, recreation, social, and sports actions account for the remainder.
Figure~\ref{fig:combined_student_plots} (right) shows the distribution of time spent across campus areas, with residences and academic buildings accounting for the majority of visits.
Overall, the agent behavior distributions provide a quantitative view of student agents' daily routines, study habits, and leisure activities, collectively depicting a colorful campus life.

\begin{figure}[!htbp]
    \centering
    \includegraphics[width=1.00\linewidth]{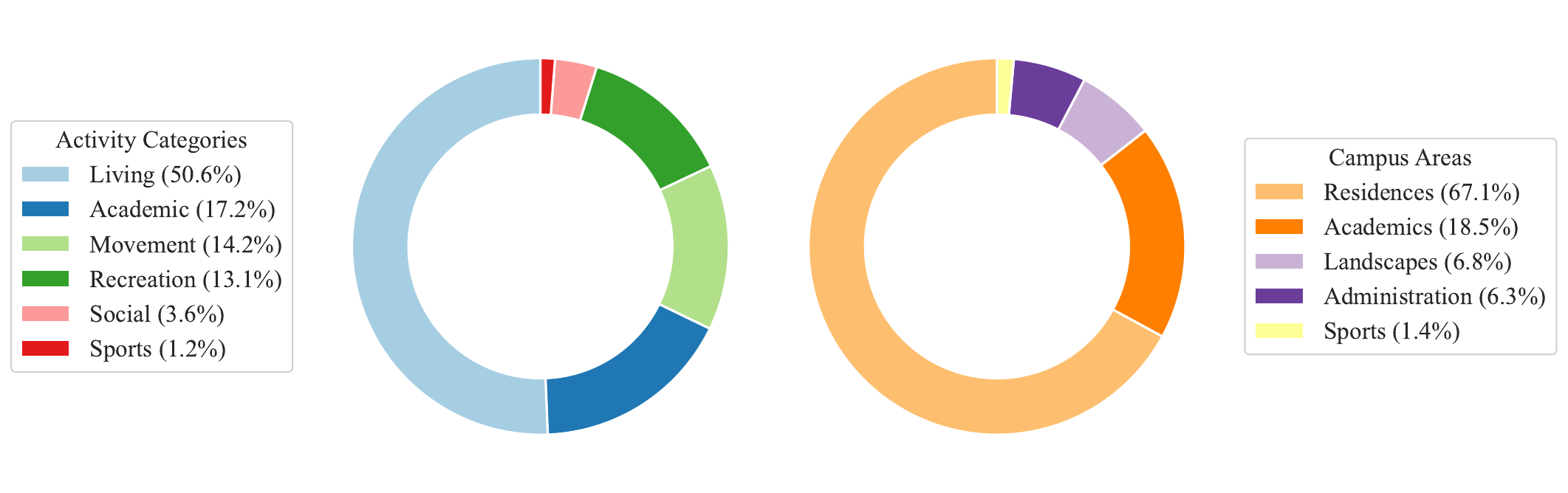}
    \caption{Distributions of daily activities (left) and time spent across campus areas (right) for student agents.}
    \label{fig:combined_student_plots}
\end{figure}

\textbf{Case Study.}
We also conduct a case study to evaluate the effectiveness of the reflection mechanism of the \plugin{Reflect Plugin}. Table~\ref{tab:dialogue_task} presents an example in which staff Agent \#101 first conducts a facility inspection with Agent \#102, identifying missing emergency supplies and drainage risks. Several ticks later, Agent \#101 draws on these earlier findings when coordinating with another staff member, Agent \#103, to ensure campus safety before a storm. The dialogues indicate that agents can recall, reflect on, and build upon memories to sustain contextually coherent, goal-directed interactions.

\begin{table}[!htbp]
\caption{Two conversations illustrating a multi-stage collaborative task in which Agent \#101 reuses earlier observations to support later collaborative work.}
\label{tab:dialogue_task}
\begin{tabularx}{\linewidth}{l X}
\toprule
\textbf{Speaker} & \textbf{Content} \\
\midrule
\multicolumn{2}{l}{\emph{An earlier conversation between Agent \#101 and Agent \#102...}} \\
\textbf{Agent \#102} & Agent \#101, have you finished the inventory? \textbf{I noticed two emergency flashlights are missing from the kit.} \\
\textbf{Agent \#101} & Got it. \textbf{I'll go to the warehouse to restock them.} Also, I noticed the \textbf{garage drainage is slightly clogged} after the rain. \\
\textbf{Agent \#102} & Please handle that. We need to prevent seepage risks before the next storm. \\
\midrule
\multicolumn{2}{l}{\emph{A conversation from several ticks later...}} \\
\textbf{Agent \#103} & Agent \#101, I am worried about the garage water accumulation. \textbf{Is the drainage system working properly?} \\
\textbf{Agent \#101} & \textbf{I found debris stuck in the channel during the re-check.} I've arranged for a pump to clear it this afternoon. \\
\textbf{Agent \#103} & That's good. And did we manage to get the supplies ready? \\
\textbf{Agent \#101} & Yes. \textbf{The missing flashlights have been replenished} and distributed to the duty rooms. We are prepared. \\
\bottomrule
\end{tabularx}
\end{table}

\section{Conclusions}

Agent-Kernel is a Multi-Agent Systems (MAS) development framework for social simulation, characterized by high adaptability, configurability, reliability, and reusability. It facilitates the modeling of heterogeneous societies, including both human and non-human entities. With sufficient computing resources, you can scale your environment to support a virtually infinite number of agents.
Furthermore, we provide a SocietyHub to facilitate sharing, content creation, and community growth. We invite you to join the Agent-Kernel community and enjoy the journey of simulation.

\bibliographystyle{IEEEtran} 
\bibliography{ref}    


\end{document}